\documentclass[letterpaper]{rsos}

\usepackage{aas_macros}
\usepackage{textcmds}

\newcommand{\ewsp}{$e\omega$-space}
\newcommand{\ldquo}{``}
\newcommand{\rdquo}{``}

\begin{document}

\title{The unexpected narrowness of eccentric debris rings: a sign of
  eccentricity during the protoplanetary disc phase}

\author{
  Grant M. Kennedy$^{1,2}$}

\address{
$^{1}$Department of Physics, University of Warwick, Gibbet Hill Road,
Coventry CV4 7AL, UK \\
$^{2}$Centre for Exoplanets and Habitability, University of Warwick, Gibbet Hill Road,
Coventry CV4 7AL, UK
}

\subject{astrophysics, extrasolar planets}


\corres{Grant Kennedy\\
\email{g.kennedy@warwick.ac.uk}}

\begin{abstract}
  This paper shows that the eccentric debris rings seen around the stars
  Fomalhaut and HD~202628 are narrower than expected in the standard
  eccentric planet perturbation scenario (sometimes referred to as
  ``pericenter glow''). The standard scenario posits an initially
  circular and narrow belt of planetesimals at semi-major axis $a$,
  whose eccentricity is increased to $e_f$ after the gas disc has
  dispersed by secular perturbations from an eccentric planet, resulting
  in a belt of width $2ae_f$. In a minor modification of this scenario,
  narrower belts can arise if the planetesimals are initially eccentric,
  which could result from earlier planet perturbations during the
  gas-rich protoplanetary disc phase. However, a primordial eccentricity
  could alternatively be caused by instabilities that increase the disc
  eccentricity, without the need for any planets. Whether these
  scenarios produce detectable eccentric rings within protoplanetary
  discs is unclear, but they nevertheless predict that narrow eccentric
  planetesimal rings should exist before the gas in protoplanetary discs
  is dispersed. PDS~70 is noted as a system hosting an asymmetric
  protoplanetary disc that may be a progenitor of eccentric debris ring
  systems.
\end{abstract}


\begin{fmtext}

\end{fmtext}


\maketitle

\section{Introduction}

Debris discs are the circumstellar discs that are seen to orbit
main-sequence stars, including our Sun. The dust that is observed
derives from a parent population of planetesimals; around other stars
this replenishment is inferred because the lifetime of the dust in the
presence of radiation and stellar wind forces is typically shorter than
the stellar age \citep[e.g.][]{1993prpl.conf.1253B}, while in the Solar
system the connection is more direct because both the dust and parent
bodies are detectable
\citep[e.g.][]{2010ApJ...713..816N,2019ApJ...881L..12P}.

In the Solar system the structure of the asteroid and Edgeworth-Kuiper
belts is shaped by planets. Indeed, these small body populations have
arguably contributed far more per unit mass to our understanding of
Solar system history than the planets themselves. Perhaps the most
famous example is the inference of Neptune's outward migration from
Pluto's capture into the 2:3 mean-motion resonance
\citep{1993Natur.365..819M}.

One key challenge for debris disc science is the successful application
of similar concepts to other stars. While a future aspiration is to
unravel the histories of other planetary systems
\citep[e.g.][]{2003ApJ...598.1321W}, one past and present goal is to
correctly infer the presence of as-yet undetectable planets via their
gravitational influence
\citep[e.g.][]{1999ApJ...527..918W,2016ApJ...827..125L}. Because these
perturbations tend to act on timescales that are longer than dust
lifetimes, the general expectation is that the planetary influence is
imprinted on the parent planetesimal population and inherited by the
collisional fragments that are observed.

Connecting disc structures, of which myriad are seen, to planets, has
proven hard, primarily because detecting the putative planets is
hard. Indeed, it is normally impossible to rule out the proposed
planets, partly because they commonly lie somewhere along a locus in
mass--semi-major axis parameter space rather than in specific locations,
but primarily because their masses can be far too small for
detection. The single successful example of prediction and subsequent
detection is for the edge-on disc $\beta$ Pictoris, whose warp was
attributed to an inclined planet \citep{1997MNRAS.292..896M} that was
subsequently discovered by direct imaging
\citep{2009A&A...493L..21L,2010Sci...329...57L}.

While $\beta$ Pic b was predicted based on the long-term ``secular''
perturbations from a misaligned planet, warps in discs are normally hard
to detect because most systems do not have the optimal edge-on
geometry. In-plane perturbations that result in azimuthally dependent
structures are more generally detectable, typically manifesting as a
significant eccentricity, which may be imaged directly or inferred from
a brightness asymmetry at lower spatial resolution. The first detection
of an eccentric debris disc was for HR~4796 \citep[which was in fact
also the second debris disc to be imaged,][the first being $\beta$
Pic]{1998ApJ...503L..79J,1998ApJ...503L..83K}, where a brightness
asymmetry (the so-called ``pericenter glow'') from low resolution
mid-infrared imaging was attributed to an unseen planet
\citep{1999ApJ...527..918W,2011A&A...526A..34M}.

Despite a high sensitivity to asymmetry (due to the exponential
dependence of flux density on dust temperature), mid-IR pericenter glow
is rarely the method by which asymmetric discs are identified, simply
because few discs are bright enough to be imaged near 10$\mu$m (both in
absolute terms, and relative to the host star). Instead, scattered light
images have proven far more successful and yielded a veritable zoo of
structures
\citep[e.g.][]{1984Sci...226.1421S,2007ApJ...671L.165H,2007ApJ...661L..85K,2015Natur.526..230B}. While
many of these images suggest the influence of unseen planets \citep[more
systems than might be surmised based on a simple expectation of
eccentric rings,][]{2016ApJ...827..125L}, a major limitation is that
these images trace small dust. This dust is subject to strong radiation
and stellar wind forces, and possibly gas drag which in addition to
opening the possibility of entirely different sculpting scenarios
\citep[e.g.][]{2009ApJ...702..318D,2019ApJ...883...68L}, makes
connecting the observed structure to the orbits of the underlying parent
planetesimals difficult. Ideally, inferences of unseen planets would be
made at longer wavelengths, where the typical grain sizes are large
enough to be immune to non-gravitational perturbations, and the observed
structure more reasonably assumed to be representative of the
planetesimal orbits.

Until recently, mm-wave observation of debris discs was limited to
photometry and marginally resolved imaging
\citep[e.g.][]{1998Natur.392..788H,2017MNRAS.470.3606H}. However, the
unprecedented sensitivity and resolution of the Atacama Large
Millimetre/Submillimetre Array (ALMA) means that well-resolved debris
disc images are now of sufficient quality to be confronted with
dynamical and collisional models. Indeed, after considering the
motivation in more detail (section \ref{s:mot}) this paper shows that
ALMA images of the debris discs in the Fomalhaut and HD~202628 systems
have a narrower radial width than is expected based on the secular
perturbation scenario originally devised to explain pericenter glow for
HR~4796 (section \ref{s:mod}). Possible origins, including the
possibility that these eccentric discs are actually a relic from the
protoplanetary disc phase, are discussed in section \ref{s:orig}, and
concluding remarks made in section \ref{s:conc}.

\section{Motivation and expectations}\label{s:mot}

The basic idea that undepins this work is that the width of any debris
ring provides information on the eccentricities of the objects that are
observed. An axisymmetric debris ring's width could be entirely
explained by orbits of a single semi-major axis $a$ and non-zero
eccentricity, as long as the pericenter angles are uniformly distributed
in azimuth. The eccentricities could however be lower, because some (or
all) of the ring width could also arise from a range of semi-major
axes. Thus, in most cases one expects to be able to derive an upper
limit on the eccentricities. Essentially the same argument applies to
eccentric debris rings, but the constraint is a bit more complex because
the pericenter angles must have a preferred direction to break the
symmetry. To understand this constraint and why it is useful, this
section briefly describes the secular perturbation mechanism that is
supposed to be the origin of eccentric debris rings \cite[see ref][for a
detailed explanation]{1999ApJ...527..918W}, and how particle
eccentricities and pericenters can be related to observable disc
structure. These ideas form the basis for the model used below.

\begin{figure}
  \begin{center}
  \includegraphics[width=0.6\textwidth]{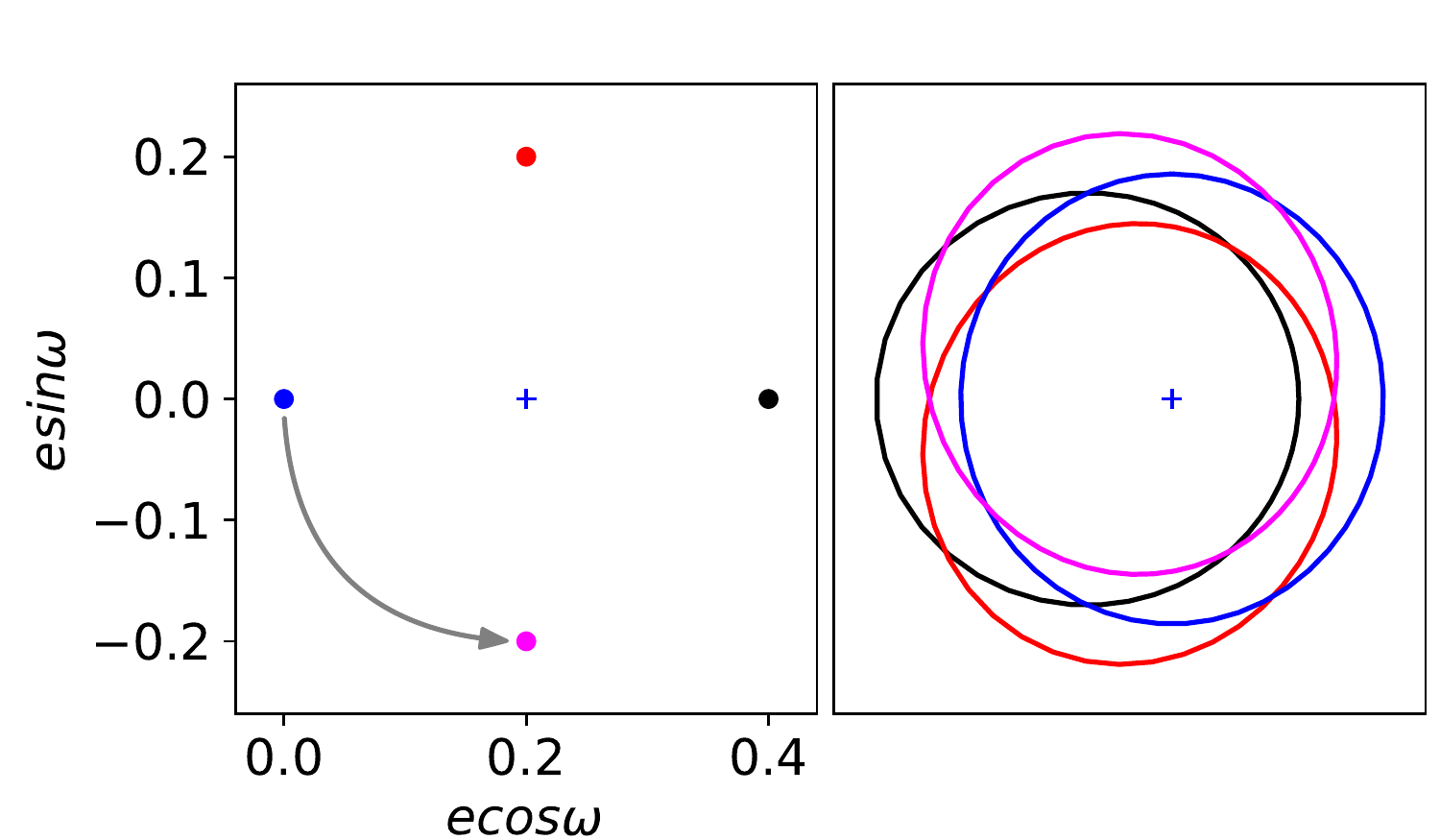}
  \caption{Illustration of how secular perturbations produce an
    eccentric debris ring, with \ewsp~in the left panel, and particle
    orbits viewed from above in the right panel. Particles that
    initially have zero eccentricity precess anticlockwise around the
    forced eccentricity (here $e_f=0.2$, $\omega=0$ is marked by the
    blue + symbol). A precessing particle starting at the blue dot
    (i.e. with $e_p=0.2$) would pass through the magenta dot, then the
    black dot, then red, then blue, etc.. The orbits for these four
    points in the precession cycle are shown in the right panel, where
    the star is marked by the blue + symbol. In particular, note that
    the black orbit is eccentric while the blue one is circular. The sum
    of many such orbits distributed in a circle around $e_f$ produces an
    eccentric ring (see upper right panel in Figure \ref{fig:eg}).}
  \label{fig:eg0}
\end{center}
\end{figure}

Secular (long-term) perturbations from an eccentric planet cause disc
particles' pericenter angles to precess, during which their
eccentricities vary in a systematic way. As applied to debris discs,
this scenario assumes that it is the large and long-lived planetesimals
that are perturbed onto eccentric orbits, and that smaller bodies
inherit these eccentric orbits when they are created. This orbital
evolution is best visualised in eccentricity--pericenter phase space
(here called \ewsp, but $hk$-space is also used), where the distance
from the origin is the eccentricity, and the anti-clockwise angle from
the positive x-axis the pericenter direction relative to some reference
direction (normally the ascending node). The planet imparts a ``forced''
eccentricity $e_f$ on test particles, whose magnitude is approximately
the planet/particle semi-major axis ratio (assuming an interior planet)
multiplied by the planet eccentricity, and whose pericenter direction is
the same as the planet's. A test particle with the forced eccentricity
stays on this orbit, but particles elsewhere in \ewsp~move continuously
in an anticlockwise circle around the forced eccentricity (i.e. their
eccentricity changes as they precess). The precession rate depends on
the strength of the planet's perturbation, via the planet mass and
planet/planetesimal semi-major axis ratio. An illustrative example for a
particle with zero initial eccentricity is given in Figure
\ref{fig:eg0}, which shows the particle and the corresponding orbits at
four points in the precession cycle. The difference between a particle's
actual eccentricity and the forced eccentricity is termed the ``free''
or ``proper'' eccentricity $e_p$, which is also the radius of the circle
drawn out by the particle. The motion in \ewsp~is such that the
particle's orbit is most eccentric only when the pericenter angle is
near zero (i.e. near the black dot); this behaviour is the key to how
secular perturbations produce an eccentric ring.

\begin{figure*}
  \includegraphics[width=0.5\textwidth]{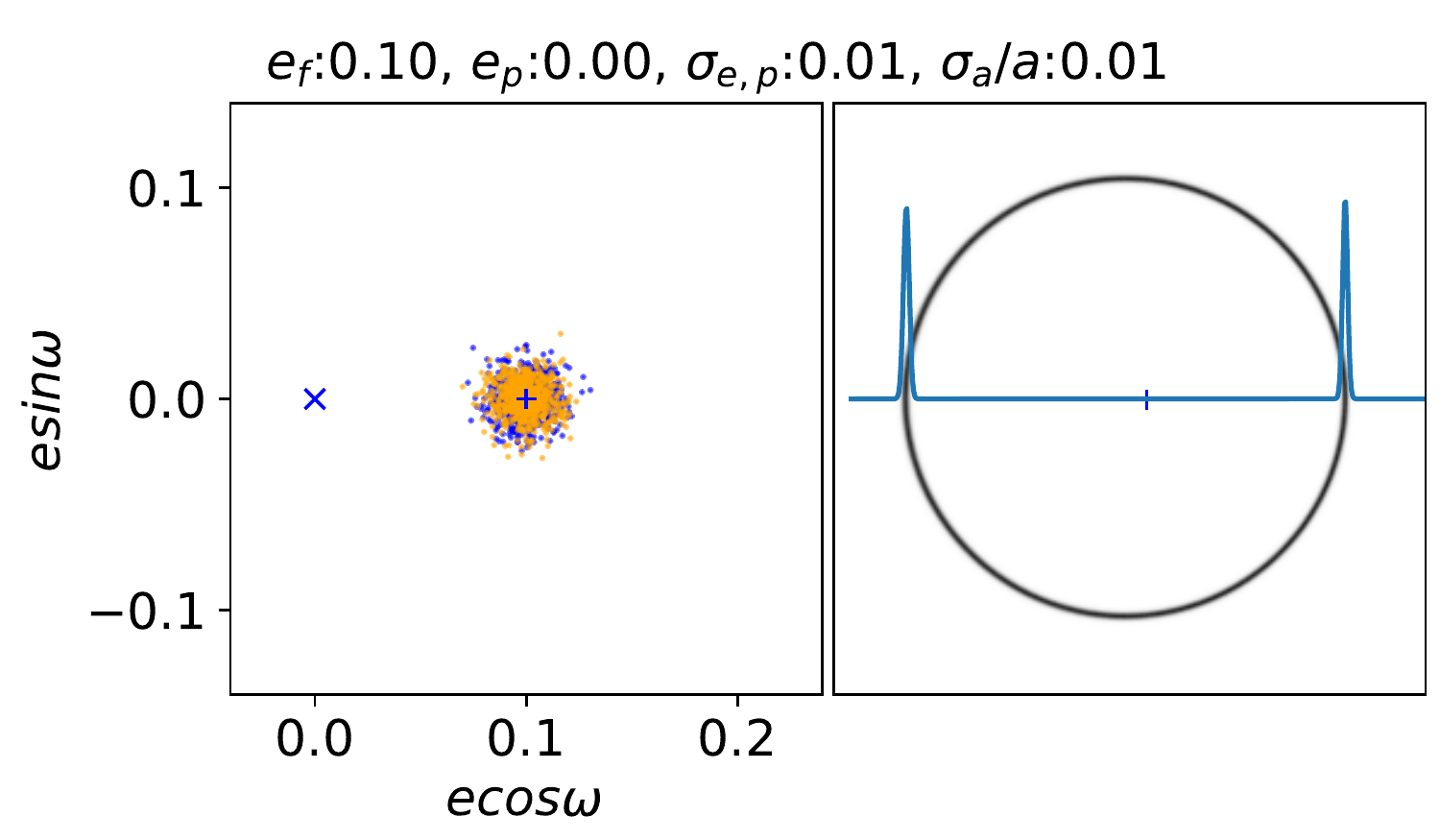}
  \includegraphics[width=0.5\textwidth]{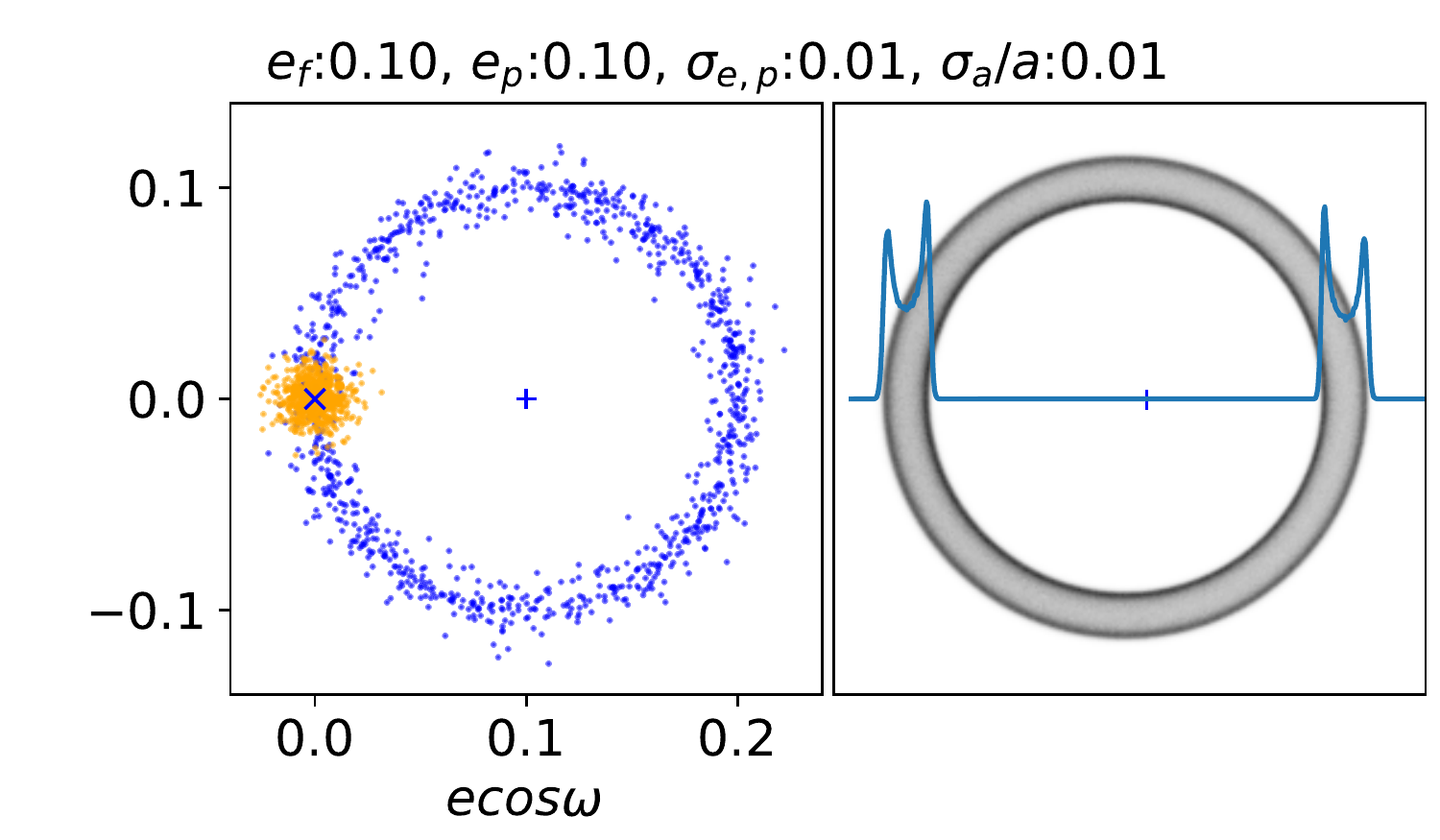}\\
  \includegraphics[width=0.5\textwidth]{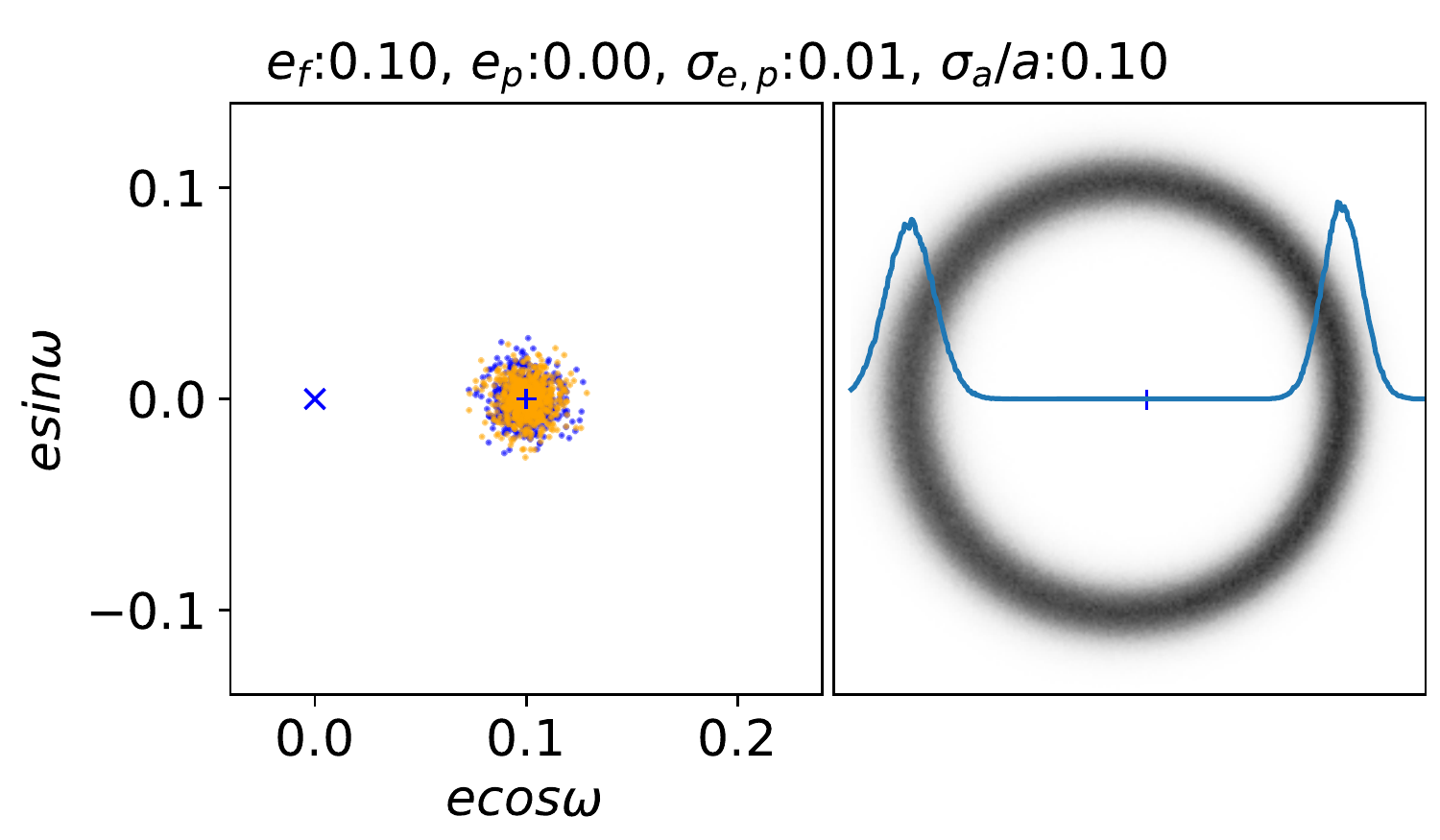}
  \includegraphics[width=0.5\textwidth]{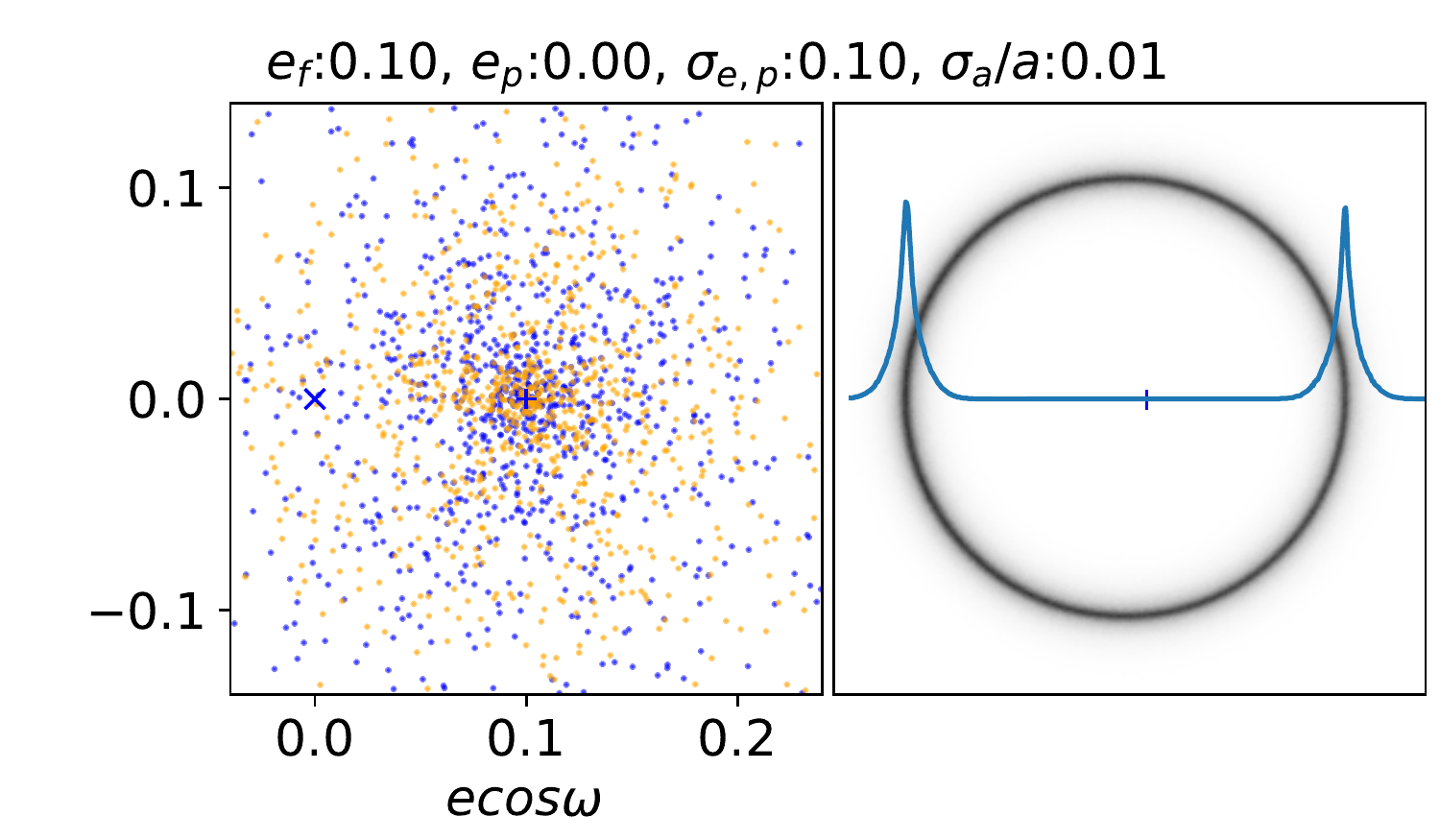}\\
  \caption{Single-planet secular perturbations with different initial
    eccentricities. In each panel the left plot shows the
    \ewsp~(initial=orange, final=blue), and the right image shows the
    resulting eccentric disc image, with the star marked by the blue +
    (assuming face-on geometry, the stated range of semi-major axes, and
    including a $1/\sqrt{r}$ weighting to account for the radially
    varying surface brightness at long wavelengths). The line in the
    right image shows the radial profile along the pericenter
    direction. The forced ecentricity is the same in all panels. The top
    left panel shows a narrow ring, while the other three show how the
    disc width can arise from small initial eccentricities (top right),
    a large initial eccentricity dispersion about the forced
    eccentricity (bottom right), or a range of semi-major axes (bottom
    left). Only a range of semi-major axes yields an azimuthally varying
    ring width.}
    \label{fig:eg}
\end{figure*}

Thus, given an initial population of particles at some common location
in \ewsp, but with a finite range of semi-major axes, the long term
evolution due to secular perturbations from an eccentric planet is that
these particles are spread around a circle of radius $e_p$ that encloses
the forced eccentricity. The time taken for particles to spread out
depends on their differential precession rates, and thus the range of
semi-major axes. The case of a Gaussian distribution of near-zero
initial eccentricities with dispersion\footnote{In this work the word
  ``dispersion'' is used to describe the width of a Gaussian
  distribution, while ``range'' refers to the full range of values
  covered by some parameter. So in the upper right panel of Figure
  \ref{fig:eg}, the initial eccentricity dispersion is small, but the
  range of final eccentricities is large. Below, both Gaussian and
  top-hat semi-major axis distributions are used, in which case
  ``range'' is used to describe the width of either distribution.}
$\sigma_{e,p}=0.01$ is shown in the upper right panel of Figure
\ref{fig:eg}, with the initial eccentricities as orange dots, and the
final eccentricities after many precession cycles as blue dots (the
semi-major axes are distributed as a narrow Gaussian with dispersion
$\sigma_a/a=0.01$). In this case, $e_p \approx e_f$, the full range of
eccentricities is
$\approx$$2e_f$, and the resulting debris ring has a uniform width of
$\approx$$2ae_f$ (looking back at Figure \ref{fig:eg0} may help make
these results clear). Alternatively, if the initial eccentricities were
very close to $e_f$, as shown in the upper left panel of Figure
\ref{fig:eg}, the final eccentricities are also near $e_f$, and the
resulting debris ring appears very narrow (i.e. approximately follows a
single eccentric orbit). Naively, one expects that the initial
planetesimal eccentricities are small, the result of damping during the
gas-rich protoplanetary disc phase, so the former of these two cases
seems more physically plausible \cite[e.g.][]{2005A&A...440..937W}. That
is, with this model one expects eccentric debris rings to have widths of
$\approx$$2ae_f$, and narrower widths would imply non-zero initial
eccentricities that are shifted towards the forced eccentricity.

Further variations on this model are of course possible; the semi-major
axis distribution can be made wide enough to dominate the belt width, in
which case the width varies as a function of azimuth (e.g. with
$\sigma_a/a=0.1$ and
$e_p=0$, see the lower left panel of Figure \ref{fig:eg}). The
dispersion of the initial eccentricity distribution can also be varied;
as
$\sigma_{e,p}$ increases the radial disc profile acquires a lower level
``halo'' that is not present when the dispersion is small (compare the
lower right and upper left panels in Figure \ref{fig:eg}). Thus,
$e_p$ changes the width, and
$\sigma_{e,p}$ the radial concentration of an eccentric ring, and the
semi-major axis distribution provides a further means to change the
width, but in a way that is azimuthally-dependent. While alternative
parameterisations could also describe an eccentric disc, it could not
achieve the same flexibility with fewer parameters, and the advantage
here is that the model is physically connected to the initial conditions
via the assumed secular perturbation scenario. Of course, whether these
parameters are all needed depends on the data in hand, which is
considered below.

\begin{figure*}
  \includegraphics[width=0.5\textwidth]{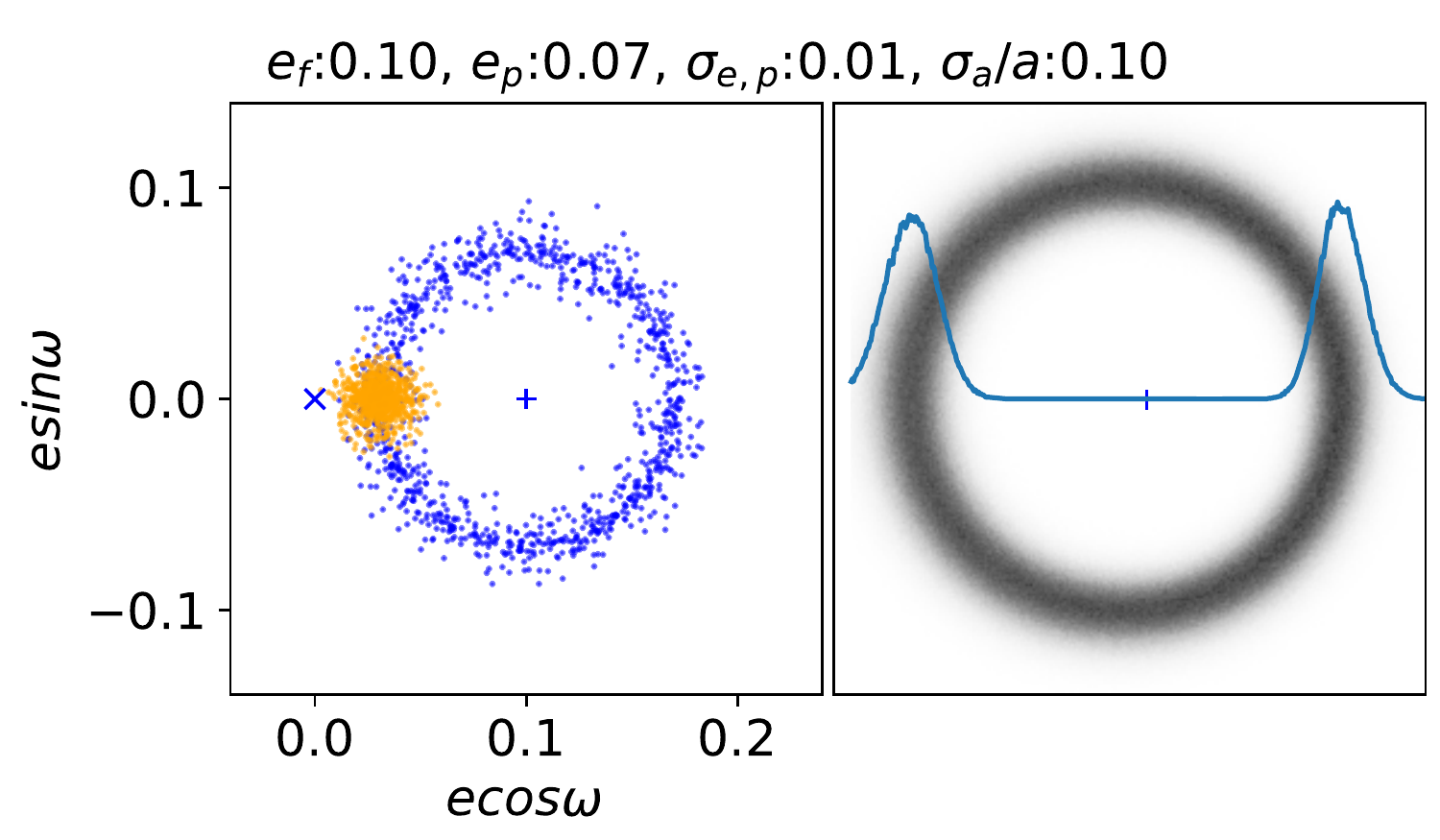}
  \includegraphics[width=0.5\textwidth]{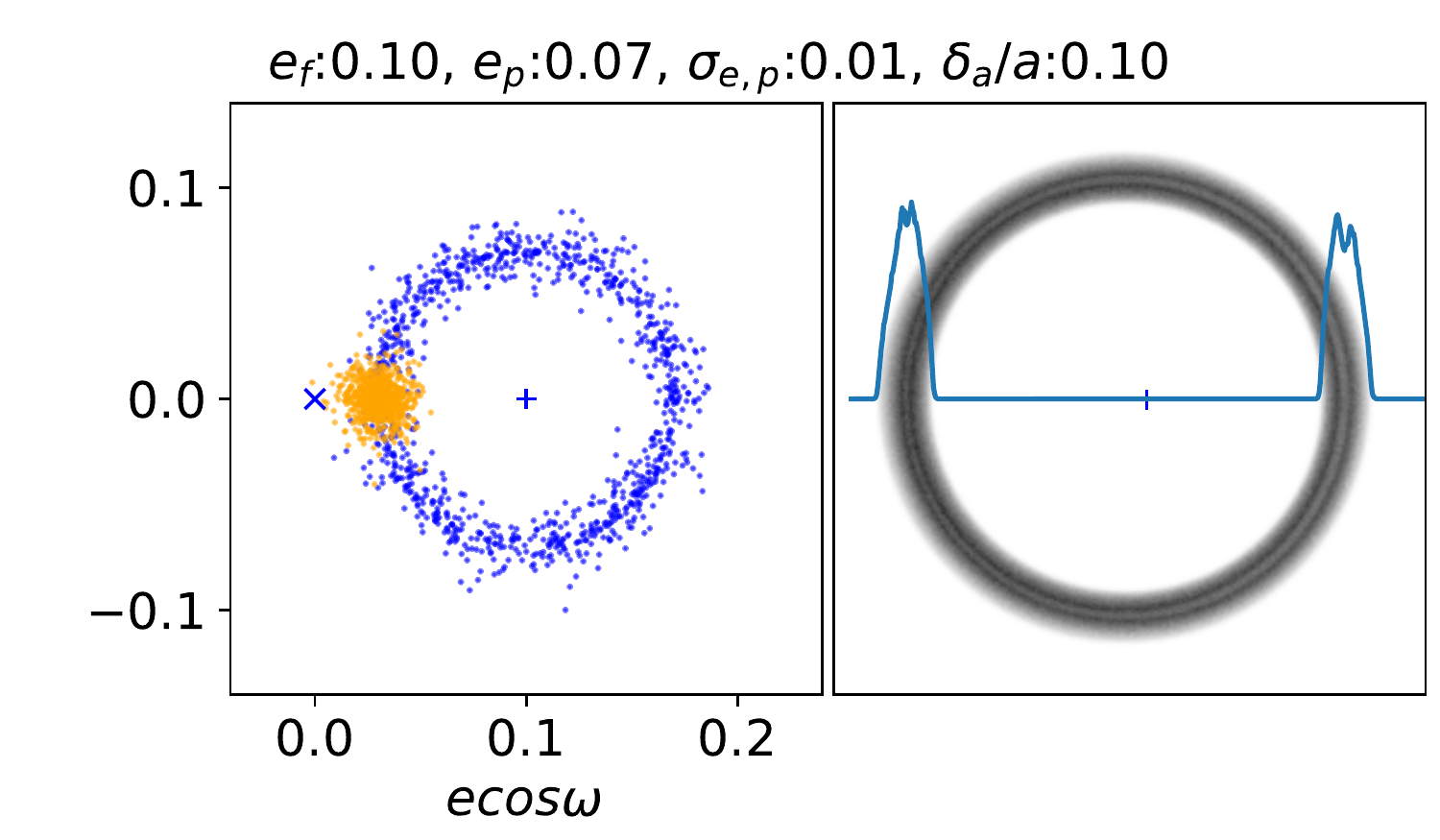}\\
  \caption{Single-planet secular perturbations with different semi-major
    axis distributions. Conventions are as in Figure \ref{fig:eg}. The
    left panel is similar to the lower left panel in Figure
    \ref{fig:eg}, but with a non-zero proper eccentricity. The right
    panel has the same eccentricity parameters, but the semi-major axis
    distribution is uniform in the range $a \pm \delta_a/2$. The
    double-ring structure in the right panel can be thought of as the
    superposition of two rings like the top right panel of Figure
    \ref{fig:eg} with semi-major axes $a-\delta_a/2$ and
    $a+\delta_a/2$. As can be seen from the radial profiles, near the
    peak brightness the disc is narrower at apocenter than pericenter
    (in contrast to all other models).}
  \label{fig:eg1}
\end{figure*}

Whether the range of semi-major axes should be small, or be distributed
in a particular way is unclear. To briefly explore this aspect, Figure
\ref{fig:eg1} shows two examples that have the same eccentricity
distributions; the left panel uses a Gaussian distribution of semi-major
axes centered at $a$ with dispersion $\sigma_a$, while the right uses a
uniform distribution between $a \pm \delta_a/2$. In terms of the
resulting disc images, the left panel in Figure \ref{fig:eg1} is similar
to the lower left panel in Figure \ref{fig:eg}, with some additional
width contributed by an increased range of eccentricities. The right
panel in Figure \ref{fig:eg1} is similar to the upper right panel in
Figure \ref{fig:eg}, but has two bright rings near the middle. These
arise because this disc is essentially a superposition of two such
rings, one at $a-\delta_a/2$ and the other at $a+\delta_a/2$. The outer
of the bright rings is the outer edge of the component with
$a-\delta_a/2$, and the inner ring the inner edge of the component with
$a+\delta_a/2$. Perhaps importantly, the distance between these rings is
not constant with azimuth, and is actually smaller near apocenter, in
contrast to all other models. Given an appropriate disc configuration,
observations that moderately resolve the disc width might be able to
detect or rule out such a variation. Note however that this example is
highly idealised, with sharp edges in the semi-major axis distribution,
and eccentricities that do not vary with semi-major axis.

\subsection{Previous work}

Lee \& Chiang \cite{2016ApJ...827..125L} developed a similar model,
which was extended to incorporate radiation forces on small grains, and
showed that many of the unusual scattered light morphologies observed
for bright debris discs can be explained by highly eccentric parent
belts ($e_f \sim 0.6$). Their work assumed a very small proper
eccentricity of 0.02, thus implicitly assuming either that the parent
belt was initially very eccentric, or that particles damp to the forced
eccentricity before they fragment to small enough sizes that they are
strongly perturbed by radiation forces. It seems unlikely that their
reproduction of scattered light structures would appear as compelling
with $e_f=e_p$, as the parent belts would be predicted to be much wider
(the effect on smaller particles is less clear, most likely the
structures would become less distinct). To take a specific example,
Esposito \emph{et al.} \cite{2016AJ....152...85E} model the HD~61005
disc and find $e_f=0.21$ and $e_p=0.08$, where the small proper
eccentricity is required to reproduce the relatively narrow parent
belt. Thus, it seems probable that if highly eccentric belts do explain
the range of scattered light structures, those belts should have proper
eccentricities that are smaller than is expected.

However, whether the range of observed debris disc structures can be
``unified'' remains unclear, as these models predict that asymmetric
structure seen in scattered light should in at least some cases be
accompanied by asymmetric structure in mm-wave observations. Few such
systems have been observed by ALMA, but HD~15115 provides a first
test. The scattered light structure is highly asymmetric and can be
explained with an eccentric belt with a pericenter direction in the sky
plane \citep{2016ApJ...827..125L}, which predicts that the mm-wave
emission should also appear asymmetric. However, MacGregor \emph{et al.}
\cite{2019ApJ...877L..32M} find that the structure is consistent with
being symmetric, suggesting that in this case a highly eccentric parent
belt is not the reason for the asymmetric scattered light structure. The
potential for such stark differences shows that connecting observations
with the underlying planetesimal belt structure is more easily done with
thermal emission at longer wavelengths, such as probed by ALMA.

\section{The ring widths of Fomalhaut and HD~202628}\label{s:mod}

\begin{figure*}
  \includegraphics[width=0.5\textwidth]{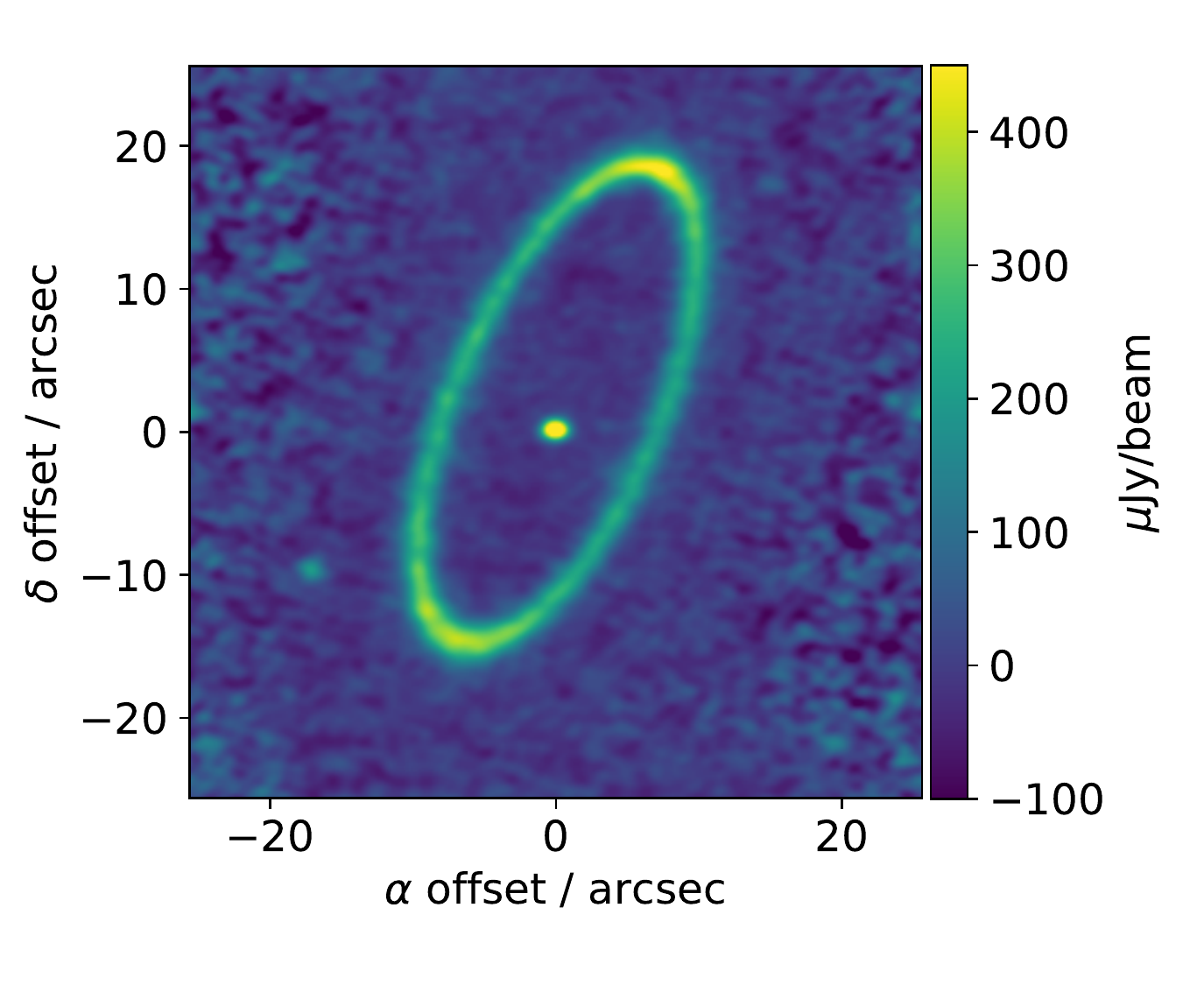}
  \includegraphics[width=0.5\textwidth]{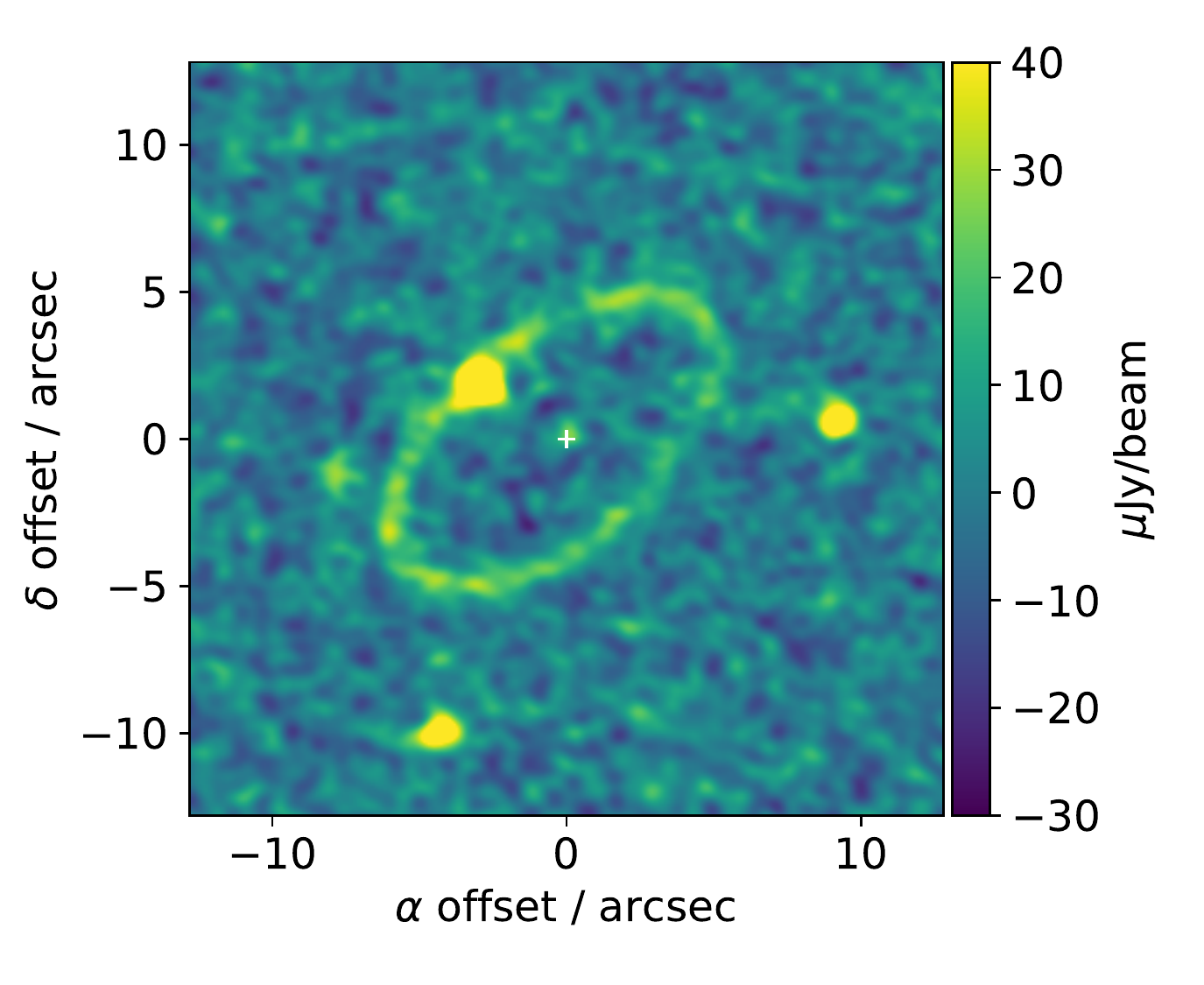}
  \caption{Naturally-weighted clean images of the Fomalhaut (left) and
    HD~202628 (right) systems.}
  \label{fig:img}
\end{figure*}

Given the motivation above, several eccentric systems are well-suited to
quantifying the ring width. Here Fomalhaut and HD~202628 are singled out
as the best cases, because their debris rings are obviously eccentric
and narrow, and because extant ALMA observations have sufficient spatial
resolution to constrain the ring width to better than the expected
$2ae_f$. That is, the null hypothesis is that the ring widths are
consistent with that expected from secular perturbations and initially
near-zero eccentricities. The assumption that the dust grains detected
with ALMA share the same orbits as the planetesimals from which they are
derived will be revisited below. Clean images of these discs are shown
in Figure \ref{fig:img}, primarily to aid the interpretation of the
residual images shown in Figures \ref{fig:fom-res} and \ref{fig:hd-res},
The Fomalhaut observations were at 1.3\,mm (band 6) and have a spatial
resolution of $1.6 \times 1.2$\,arcsec resolution ($12 \times 9$\,au at
the 7.7\,pc distance of Fomalhaut). The large angular size relative to
the primary beam means that seven pointings were required to cover the
disc adequately. The HD~202628 observations were also at 1.3\,mm and use
a single pointing. The spatial resolution is $0.9 \times 0.8$\,arcsec,
which is $21 \times 19$\,au at the 23.8\,pc distance of HD~202628. The
reader is referred to the papers cited below for more observational
details.

The HD~202628 ring was modelled by Faramaz \emph{et al.}
\cite{2019AJ....158..162F}, and with the assumption of constant surface
density (justified by a marginally radially resolved image) the width
was found to be 22\,au. Given their best-fit eccentricity of 0.09 the
expected width of $2ae_f = 28$\,au is larger than observed, suggesting
that the proper eccentricity is lower than the forced eccentricity. The
ring is at best marginally resolved radially and not detected with a
high signal to noise ratio, but the star is detected at the expected
location which significantly improves the constraints on the ring
geometry. Their model was simplified in that it used an offset circular
ring; here the data are modelled again using the more complex eccentric
ring model. Given the data this is not necessarily a superior approach,
but has the benefit that ring parameters directly related to the secular
perturbation scenario are derived explicitly, and their (joint)
posterior distributions quantified.

\begin{figure}
  \begin{center}
  \includegraphics[width=0.6\textwidth]{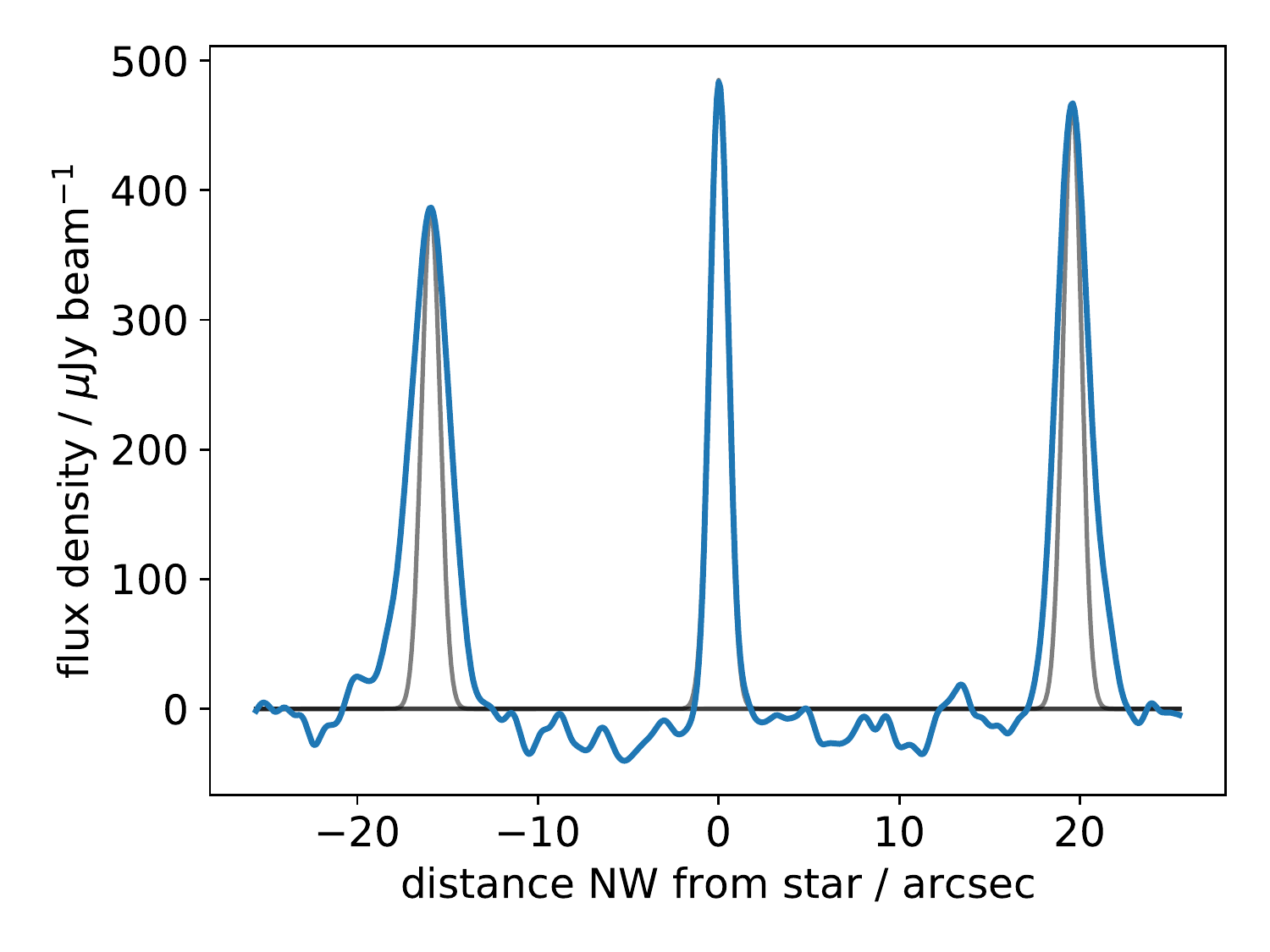}
  \caption{Radial surface brightness profile along the disc major axis
    for Fomalhaut. The blue line shows the data, and the grey lines show
    Gaussian profiles with FWHM 1.2\,arcsec (i.e. the spatial resolution)
    scaled to each peak. The measured (unconvolved) disc FWHM in each
    ansa are 2.6 (SE) and 2.1 (NW)\,arcsec, or 20 and 16\,au.}
  \label{fig:fom_rprof}
\end{center}
\end{figure}

The width of the Fomalhaut ring has also been suggested to be narrower
than expected by White \emph{et al.} \cite{2017MNRAS.466.4201W} and
MacGregor \emph{et al.} \cite{2017ApJ...842....8M}, with these works
finding full-width at half-maxima (FWHM) of 13-13.5\,au. This width is to
be compared to the expectation of $2ae_f = 33$\,au. However, the
significance of this narrowness is not particularly clear for two
reasons. Figure 5 of MacGregor \emph{et al.} \cite{2017ApJ...842....8M}
suggests that the disc has a FWHM along the major axis of
$\sim$5\,arcsec (39\,au), but the contours in their Figure 1 show that the
disc is at most only $\sim$2 beams (i.e. $\sim$2\,arcsec) wide where it
is detected (suggesting an axis labelling/scaling error). Figure
\ref{fig:fom_rprof} shows the radial profile for a naturally weighted
image with a 1\,arcsec wide swath through the star and along the disc
major axis. The disc widths along the disc major axis are 16 and 20\,au,
with the NW ansa being narrower, and accounting for the beam width of
about 1.2\,arcsec these widths are consistent with a true disc FWHM of
about 13\,au (i.e. much less than 33\,au).

A second issue is that the modelling in MacGregor \emph{et al.}
\cite{2017ApJ...842....8M} finds a proper eccentricity ($0.06 \pm 0.04$)
that is consistent with the forced eccentricity (which was
$0.12 \pm 0.01$). This result therefore suggests that the disc can be
consistent with a width of 33\,au (4.3\,arcsec) at 2$\sigma$, which is
clearly incompatible with the data (e.g. from Figure \ref{fig:fom_rprof}
here, or the contours in their Figure 1). The probable reason lies with
the details of the modelling, which generated disc images by generating
N particles on eccentric orbits (i.e. as in Figures \ref{fig:eg} and
\ref{fig:eg1}). In that work $N= 10^4$ particles were used, but in the
course of this work at least 10$^6$ particles were found to be
necessary. With small $N$, shot noise renders two model images with
identical parameters sufficiently different that their $\chi^2$ can also
be very different. This leads to issues with convergence of model
fitting, including low acceptance fractions in Markov chain Monte Carlo
(MCMC) fitting, and inflated uncertainties. This issue appears to have
affected the uncertainties more than the best-fit parameters (which
largely agree with the results here), although some differences are
discussed below.

Finally, MacGregor \emph{et al.} \cite{2017ApJ...842....8M} used a
uniform range of semi-major axes and in fact yields the double
inner-ring structure discussed above, which can be seen (albeit
indistinctly) in their Figure 2. The radial profile in Figure
\ref{fig:fom_rprof} does indeed show that the disc is narrower towards
the NW ansa (which is near the apocenter), and whether this model
provides a good explanation for the data is explored below.

\subsection{The model}\label{ss:model}

A slightly more complex version of the secular perturbation model used
by MacGregor \emph{et al.} \cite{2017ApJ...842....8M} was used to model
the ALMA data for Fomalhaut and HD~202628. Bascially, $N$ particles are
generated that sample distributions of eccentricity, inclination,
semi-major axis, and true anomaly. These are then binned into an 2d
array for the desired viewing geometry to create an image. The
fundamental assumption is that the particles have evolved to their
current state due to secular perturbations (i.e. are spread evenly
around the forced pericenter in \ewsp). This assumption means that the
models are created analytically, with no need for $n$-body
simulations. From this assumption, and similar secular timescales for
inclination evolution, it also follows that the particle nodes are also
spread randomly, meaning that any initial misalignment ($i_f$) of the
particles with the planet's orbital plane causes the disc to have a
height $2i_f$. Viewed edge-on the disc appears brighter at the upper and
lower surfaces, because their sinusoidal vertical motions means that
particles spend more time there (i.e. similar to the limb-brightening
effect seen in the top right panel of Figure \ref{fig:eg}). While the
modelling here does not resolve the scale height of either disc, a
finite vertical extent is included to include any effect on the other
parameters.

The model parameters are as follows: i) $x_0$ and $y_0$ allow for any
shift of the star+disc position from the observation phase center, ii)
sky geometry position angle $\Omega$, inclination $i$, and forced
pericenter angle $\omega$ (measured from $\Omega$), iii) total disc and
star flux densities $F$ and $F_{\rm star}$, iv) disc semi-major axis
$a_0$ and Gaussian width $\sigma_a$ or full width $\delta_a$, v) forced
eccentricity $e_f$, proper eccentricity $e_p$ (radius of the circle in
\ewsp), and proper eccentricity Gaussian dispersion $\sigma_{e,p}$, vi)
forced inclination $i_f$ and Gaussian inclination dispersion
$\sigma_{i,p}$, and vii) a data re-weighting parameter $f_w$ (see
below). A total of 15 basic parameters are included, and three more are
included in the model for HD~202628 for the position and flux of a
bright point source that could otherwise influence the model
results. The important difference compared to the model used by
MacGregor \emph{et al.} \cite{2017ApJ...842....8M} is the inclusion of a
proper eccentricity dispersion, which as shown in Figure \ref{fig:eg}
provides an additional means to introduce a finite disc width. Given
that the Fomalhaut ring appears narrower at apocenter than pericenter,
thus disfavouring a range of semi-major axes, alternative means of
generating the disc width are potentially important.

Practically, model images are created with the following steps: i) the
eccentricity parameters $e_p$ and $\sigma_{e,p}$ are used to create $N$
initial particles in a 2-d Gaussian in \ewsp. As these will be precessed
into a circle around $e_f$, $e_p$ is taken to be the distance from $e_f$
towards the origin in \ewsp~(these are the orange dots in Figures
\ref{fig:eg} and \ref{fig:eg1}), ii) these particles are then
distributed evenly around the forced eccentricity, giving a set of $N$
eccentricities and pericenters (i.e. the blue dots in the same figures),
iii) $N$ inclinations are generated with a Gaussian distribution of
dispersion $\sigma_{i,p}$ centered on $i_f$, and these are given
randomly chosen ascending nodes, iv) $N$ random mean anomalies are
generated, and using the eccentricity of each particle these are
converted into true anomalies, v) using the eccentricities and true
anomalies, the radius of each particle from the star is calculated, and
using the true anomaly, pericenter, and node, the angular particle
locations are calculated. These steps generate an initial 3-d particle
model of the disc viewed from above, with the forced pericenter measured
anticlockwise from the positive $x$ direction. The final steps are vi)
to apply two rotations to move these particles to the observed geometry;
first the $y$ coordinates are multiplied by $\cos{i}$ to incline the
disc, and then the $x,y$ coordinates are rotated by $\Omega+\pi$ to
place the ascending node at the observed position angle, and vii)
finally, the image is generated by binning all particles into a 2-d
grid, including a $1/\sqrt{r}$ weighting to account for decreasing
temperature with radius.

Model images computed using $N=10^7$ are compared with the ALMA
visibilities using the \texttt{GALARIO} \citep{2018MNRAS.476.4527T}
software (which returns a $\chi^2$ value for a given model
image). Absolute uncertainties are first estimated with the CASA
\texttt{statwt} task, which derives weights ($=1/\sigma^2$) based on the
variance of the visibilities. The parameter space is explored using
\texttt{MultiNest} \citep{2009MNRAS.398.1601F} and \texttt{emcee}
\citep{2013PASP..125..306F}, with the log likelihood of a given model:
\begin{equation}\label{eq:like}
  \ln \mathcal{L} = -\frac{1}{2} \left( \chi^2 f_w + \sum_i^M 2 \ln \frac{2 \pi}{w_i
      f_w} \right) \, ,
\end{equation}
where there are $M$ complex visibilities $V$, with weights
$w$\footnote{The factor $f_w$, which the visibility weights are
  multiplied by, is included because CASA \texttt{statwt} is not
  guaranteed to produce accurate absolute uncertainties. The log
  likelihood function used in the fitting is therefore derived by
  assuming that the data are distributed as a Gaussian about the model
  with the correct normalisation. The factor two in the summation of
  equation (\ref{eq:like}) arises because each visibility is assumed to
  have two independent degrees of freedom (i.e. amplitude and
  phase).}. While both methods give similar results, the potential
advantage of using \texttt{MultiNest} is that it explores \emph{all}
parameter space within given ranges, meaning that similarly well-fitting
solutions in the parameter space can be identified (though no
multi-modal solutions were found here). Only the Multinest results are
reported here, which used 75 live points to search a broad parameter
space around previously found best-fit parameters. All parameters have
uniform priors in linear space.

\subsection{Fomalhaut}

The same seven-pointing data of the Fomalhaut disc that were presented
in MacGregor \emph{et al.} \cite{2017ApJ...842....8M} were modelled as
described above. The data were calibrated using the observatory-supplied
script, with the only additional steps being spectral averaging to a
single channel per spectral window (spw), 30-second time averaging, and
re-weighting of all visibilities using the CASA~5.6 \texttt{statwt}
task. The signal to noise ratio of the disc detection is sufficiently
high and the disc sufficiently large that the finite ALMA bandwidth must
be accounted for (i.e. the baselines in $uv$ space vary sufficiently
across channels that a single frequency cannot be assumed across all
channels). The $\chi^2$ are therefore computed using visibilities that
assume the average frequency for each spectral window, for each of the
seven pointings, and the 28 $\chi^2$ values summed. The precision of the
pointings was found to be good enough relative to the resolution that no
per-pointing offset parameters were necessary.

\begin{figure*}
  \includegraphics[width=0.5\textwidth]{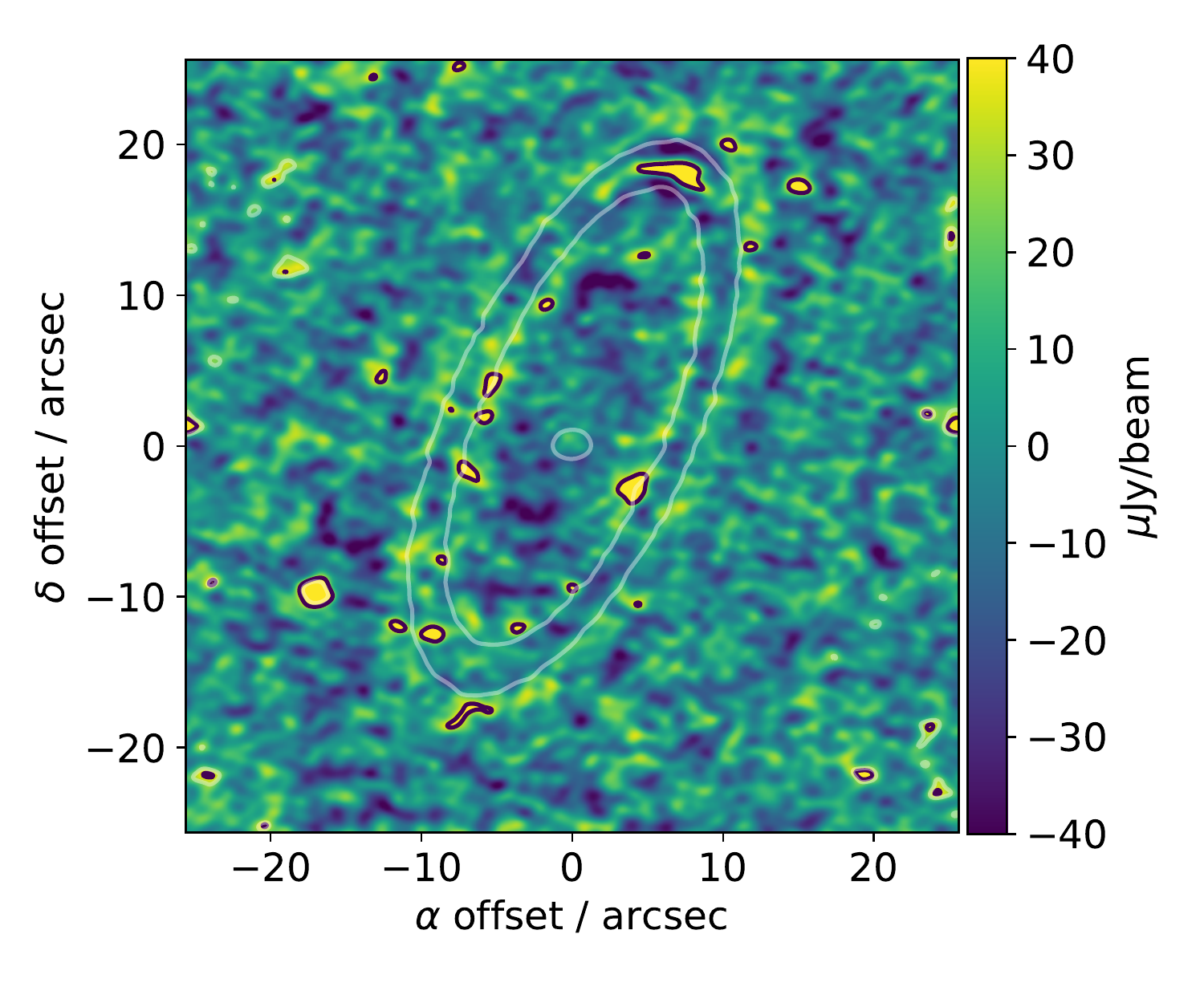}
  \includegraphics[width=0.5\textwidth]{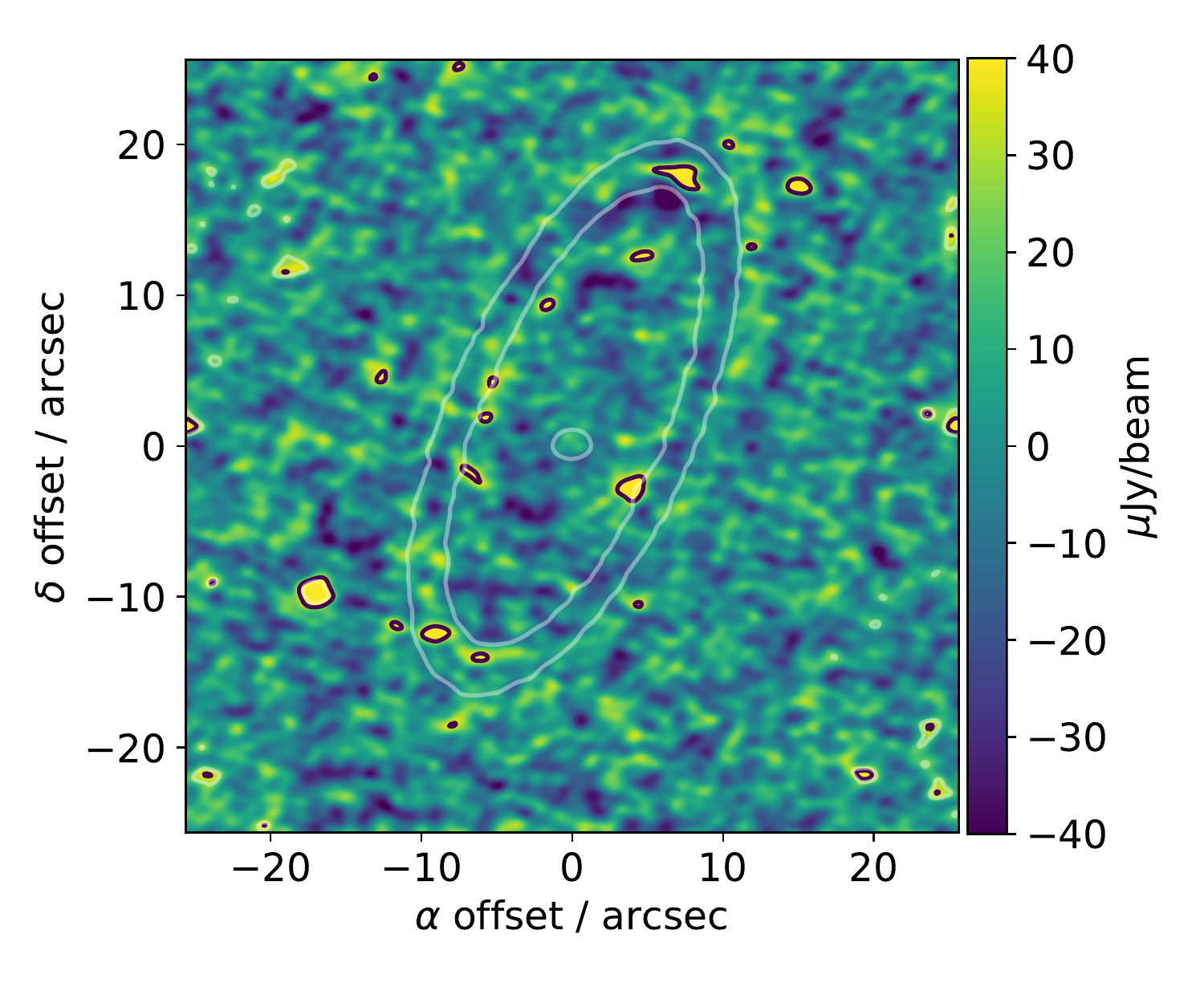}
  \caption{Residual maps for the best-fit Uniform simple (left) and full
    (right) models of the Fomalhaut ring. The white contour shows the
    location of the ring, and black contours highlight parts of the
    image above 3 times the noise level of 13$\mu$Jy\,beam$^{-1}$.}
  \label{fig:fom-res}
\end{figure*}

Four slightly different models were fitted to the Fomalhaut data; two
``full'' models that use all parameters noted above, with ``Gaussian''
and ``Uniform'' semi-major axis distributions, and two ``simple'' models
with the same two semi-major axis distributions, and where the
eccentricity and inclination dispersion parameters, and the forced
inclination, are set near zero (i.e. vertically flat models that look
similar to the upper right or lower left panels in Figure
\ref{fig:eg}). From the radial profile in Figure \ref{fig:fom_rprof} it
is clear that a purely Gaussian radial distribution is a poor
description of the data, and this is confirmed by a poor fit with the
Gaussian simple model, so this model is not discussed further. The
Gaussian full model however yields a very good fit, because the
additional radial extent allowed by the eccentricity dispersion can
account for surface brightness interior and exterior to the brightest
part of the ring. Both Uniform models provide good fits, though the full
model does a much better job of reproducing the low surface brightness
immediately interior and exterior to the ring. A test using the Schwarz
criterion\footnote{Also known as the Bayesian Information Criterion,
  ${\rm BIC} = \chi^2+k \ln n$, where k is the number of parameters and
  n the degrees of freedom (here $k$ is 12 or 15 for the simple and full
  models, and $n \approx 1.7 \times 10^6$ for Fomalhaut). Smaller BIC
  numbers are better, with a decrease of more than about 6 indicating
  that additional parameters are warranted given the data.}
\cite{1978AnSta...6..461S} finds that while the second ``penalty'' term
increases by 45 for the full models relative to the simple uniform
model, the $\chi^2$ decreases by 240. Such a large decrease indicates
that for Fomalhaut the inclusion of these extra parameters is justified
by the improvement in the fit.

\begin{table*}
  \begin{center}
    \begin{tabular}{lccc}
                  & \multicolumn{3}{c}{Fomalhaut} \\
      & Gaussian full & Uniform full & Uniform simple \\
\hline
      $a_0$ & $18.17 \pm 0.04$ & $18.18 \pm 0.05$ & $18.19 \pm 0.01$ \\
      $\sigma_a$ or $\delta_a$ & $<0.28$ & $<1.00$ & $2.09 \pm 0.06$ \\
      $e_f$ & $0.125 \pm 0.001$ & $0.125 \pm 0.001$ & $0.126 \pm 0.001$ \\
      $e_p$ & $0.019 \pm 0.004$ & $0.019 \pm 0.004$ & $0.051 \pm 0.001$ \\
      $\sigma_{e,p}$ & $0.090 \pm 0.004$ & $0.089 \pm 0.002$ & -- \\
      $i_f$ & $<0.19$ & $<0.22$ & -- \\
      $\sigma_{i,p}$ & $<0.11$ & $<0.12$ & -- \\
      \hline
      \\
      & \multicolumn{3}{c}{HD~202628} \\
      & Gaussian full & Uniform full & Uniform simple \\
      \hline
      $a_0$ & $6.6 \pm 0.08$ & $6.55 \pm 0.08$ & $6.46 \pm 0.04$ \\
      $\sigma_a$ or $\delta_a$ & $<0.4$ & $<1.2$ & $<1.3$ \\
      $e_f$ & $0.12 \pm 0.02$ & $0.12 \pm 0.02$ & $0.12 \pm 0.01$ \\
      $e_p$ & $<0.08$ & $<0.08$ & $<0.08$ \\
      $\sigma_{e,p}$ & $<0.11$ & $<0.12$ & -- \\
      $i_f$ & $<0.5$ & $<0.5$ & -- \\
      $\sigma_{i,p}$ & $<0.4$ & $<0.4$ & -- \\
      \hline
    \end{tabular}\caption{Partial list of best fit model parameters,
      with the ring radius and width in units of arcseconds and
      inclinations in radians (full posterior parameter distributions
      are shown in Figures
      \ref{fig:fom-da-full}--\ref{fig:hd-g-full}). Uncertainties are
      1$\sigma$, and upper limits are 3$\sigma$.}\label{tab:res}
  \end{center}
\end{table*}

The modelling results are summarised in Table \ref{tab:res} and Figure
\ref{fig:fom-res} which show the (dirty) residual map for the Uniform
models (the Gaussian full model residuals are indistinguishable to the
Uniform full model, Figures \ref{fig:fom-da-full}--\ref{fig:fom-g-full}
show the posterior parameter distributions for most parameters).  All
models find the same disc position angle of
$\Omega = 156.4 \pm 0.1^\circ$ and inclination $i = 66.6 \pm
0.1^\circ$. Figure \ref{fig:fom-res} shows that the models are a good
representation of the data, with the only large residual in the NW
ansa. This residual noise lies within the ring and has negative
residuals interior and exterior, suggesting that the ring is narrower
than the model at this location.

The Uniform simple model, which is essentially the same as that used by
MacGregor \emph{et al.} \cite{2017ApJ...842....8M}, finds very similar
results, but with the expected smaller uncertainties. All models find
that the pericenter location is located $41 \pm 1^\circ$ anti-clockwise
from the SE disc ansa (i.e. the ascending node). This value is most
likely different to the $23 \pm 4^\circ$ found by MacGregor \emph{et
  al.}  \cite{2017ApJ...842....8M} for the reasons discussed
earlier.

\begin{figure}
  \begin{center}
  \includegraphics[width=0.6\textwidth]{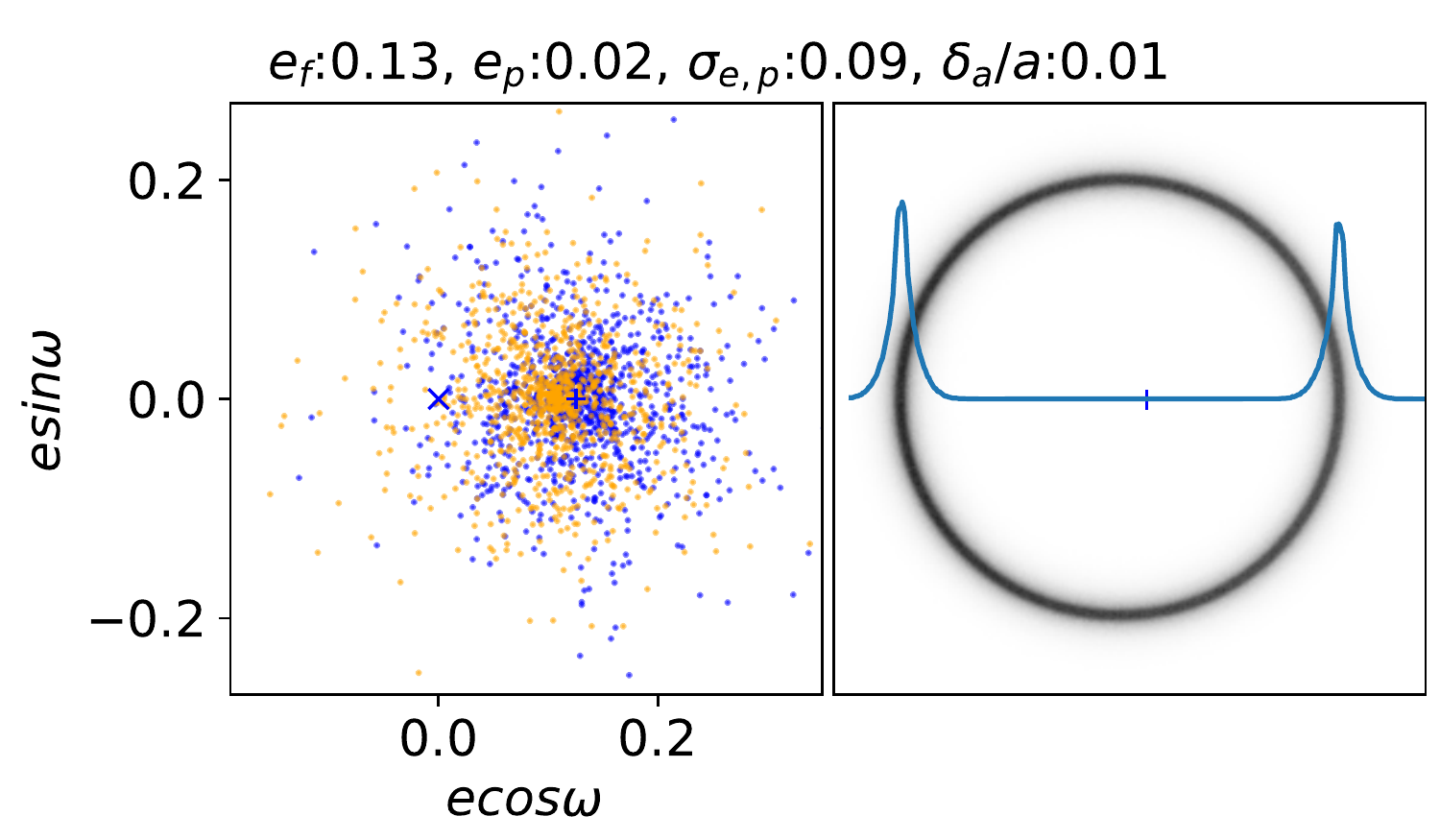}
  \caption{Best-fit Uniform full model to Fomalhaut, shown face-on with
    pericenter along the positive x axis. Conventions are as in Figure
    \ref{fig:eg}. The ring has a narrow bright component whose width is
    set by $e_p$, and is surrounded by a lower surface brightness
    ``halo'' whose width is set by $\sigma_{e,p}$.}
  \label{fig:fom-ecc}
\end{center}
\end{figure}

Less obvious from Figure \ref{fig:fom-res} is that a greater
eccentricity dispersion via the $\sigma_{e,p}$ parameter improves the
fit; low level ($\sim$1$\sigma$) residuals just interior/exterior to the
brightest part of the belt remain for the Uniform simple model (left
panel), but are reduced for the full models (right panel). The narrower
disc width near apocenter means that narrow semi-major axis
distributions are preferred, and thus for the full models the disc width
is instead generated by a combination of non-zero proper eccentricity
and eccentricity dispersion. The best-fit model is shown at high
resolution with the \ewsp~in Figure \ref{fig:fom-ecc}, which shows that
the model is composed of a narrow top-hat-like ``core'' that is produced
by the relatively small typical proper eccentricity $e_p$, and a
``halo'' that is produced by the dispersion in proper eccentricity
$\sigma_{e,p}$. The semi-major axis distribution does not contribute to
the structure, so the Gaussian full model looks the same.

While it was suggested earlier than the Uniform model may be better able
to produce a disc that is narrower near apocenter through a quirk of
combining a range of semi-major axes and proper eccentricities, this
possibility is not borne out by the modelling, primarily because the
best models do not require a range of semi-major axes.

Overall, these models show that the disc is consistent with an eccentric
ring model, but only if the proper eccentricities are smaller than the
forced eccentricity. While the constraints on the model parameters
appear very small, this is at least in part because of the restricted
nature of the models themselves, which for example are relatively
inflexible in terms of semi-major axis distributions. While the
conclusion of small proper eccentricities is robust from both empirical
and modelling approaches, further insight into the azimuthal dependence
of the ring width can only be gained with higher resolution imaging.

\subsection{HD~202628}

The same modelling procedure was applied to HD~202628. The 12m array
data from Faramaz \emph{et al.} \cite{2019AJ....158..162F} were obtained
from the ALMA archive and calibrated using the observatory-supplied
script (the ACA data have insufficient resolution to be valuable here,
and are dominated by a background source, so were not included). The
CASA \texttt{statwt} command was used to reweight the data, and all
visibilities were exported and modelled assuming a single
wavelength. Compared to Fomalhaut this disc is smaller and detected at
lower s/n, so the assumption of a single wavelength was found to be
adequate. The same four models were fitted, and all four are able to
reproduce the data well and have similar $\chi^2$ values (i.e. including
the additional parameters in the full models is not well justified given
the data).

\begin{figure}
  \begin{center}
  \includegraphics[width=0.6\textwidth]{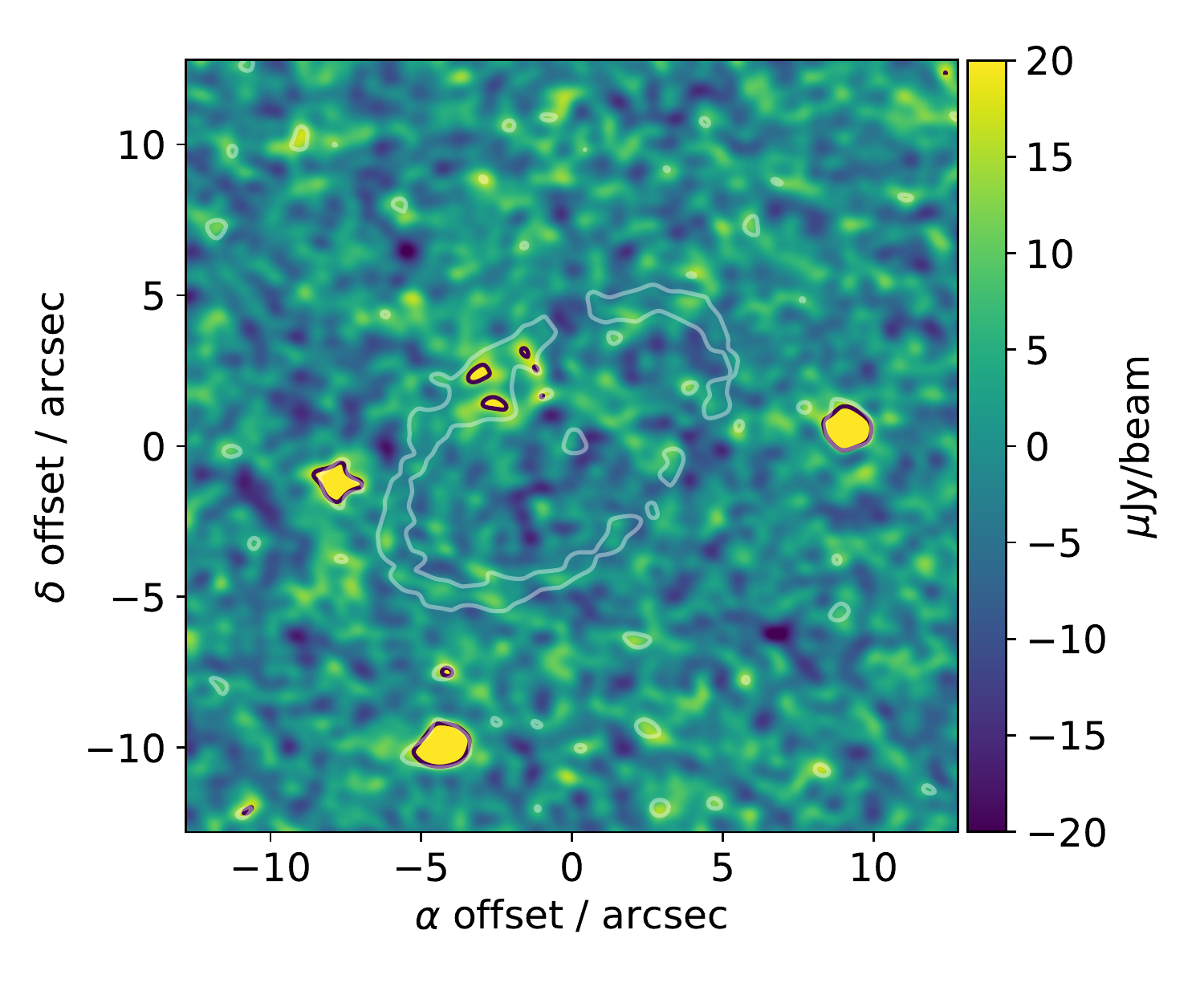}
  \caption{Residual map for the best-fit Gaussian full model of the
    HD~202628 ring. The white contour shows the location of the ring,
    and black contours highlight parts of the image above 3 times the
    noise level of 6$\mu$Jy\,beam$^{-1}$. A bright point source has been
    subtracted on the NE part of the disc, while three others were not
    included in the model. All are assumed to be unrelated to the disc.}
  \label{fig:hd-res}
\end{center}
\end{figure}

The results of the modelling are summarised in Table \ref{tab:res}, and
Figures \ref{fig:hd-res}, which shows the (dirty) residual map for the
Gaussian full model (Figures \ref{fig:hd-da-full}--\ref{fig:hd-g-full}
show the posterior parameter distributions). All models find the same
disc position angle of $\Omega = 130.7 \pm 0.1^\circ$ and inclination
$i = 57.4 \pm 0.1^\circ$. The best-fit parameters are in general
agreement with Faramaz \emph{et al.}  \cite{2019AJ....158..162F}, though
with a slightly forced larger eccentricity (0.12 here vs. their
0.09). Whether the pericenter directions agree is unclear; their Figure
3 suggests that pericenter lies near the NW ansa, but Figure 10 puts it
more directly West of the stellar position. The latter appears
consistent with the value of 143$^\circ$ from the SE ansa found here so
seems more likely to be correct. These differences may be related to the
difference in eccentricity, and will be addressed in the near future
(Faramaz et al., in preparation).

As with Fomalhaut, the residuals show that the model is a good
representation of the data, though achieving this is less of a challenge
for HD~202628 given the much lower s/n. No significant residuals that
are clearly related to the disc remain, though imperfect subtraction of
the bright point source NW of the star is apparent. While the
eccentricity and disc width parameters $\sigma_a$ and $e_p$ are given as
upper limits, this is based on their 1-dimensional posterior
distributions; inspection of the 2-d distributions finds that these two
parameters cannot simultaneously be zero, implying that the ring is
radially resolved (albeit marginally).

As with Fomalhaut, the best fit model finds that the proper eccentricity
is significantly lower than the forced eccentricity,\footnote{This
  significance is better verified directly from the posterior parameter
  distributions in the appendix, as the values in Table \ref{tab:res} do
  not reflect asymmetric uncertainty ranges, nor that the 3$\sigma$
  limit for a parameter is not necessarily 3 times the 1$\sigma$
  uncertainty.} again implying that the ring cannot be modelled as a
secularly perturbed ring whose eccentricities were initally near
zero. The lower s/n and spatial resolution means that the constraints on
the parameters are not as strong, and higher spatial resolution data
would be needed to obtain any insight into how the ring width varies as
a function of azimuth.

\section{Eccentric ring origins - why so narrow?}\label{s:orig}

\begin{figure}
  \begin{center}
  \includegraphics[width=0.6\textwidth]{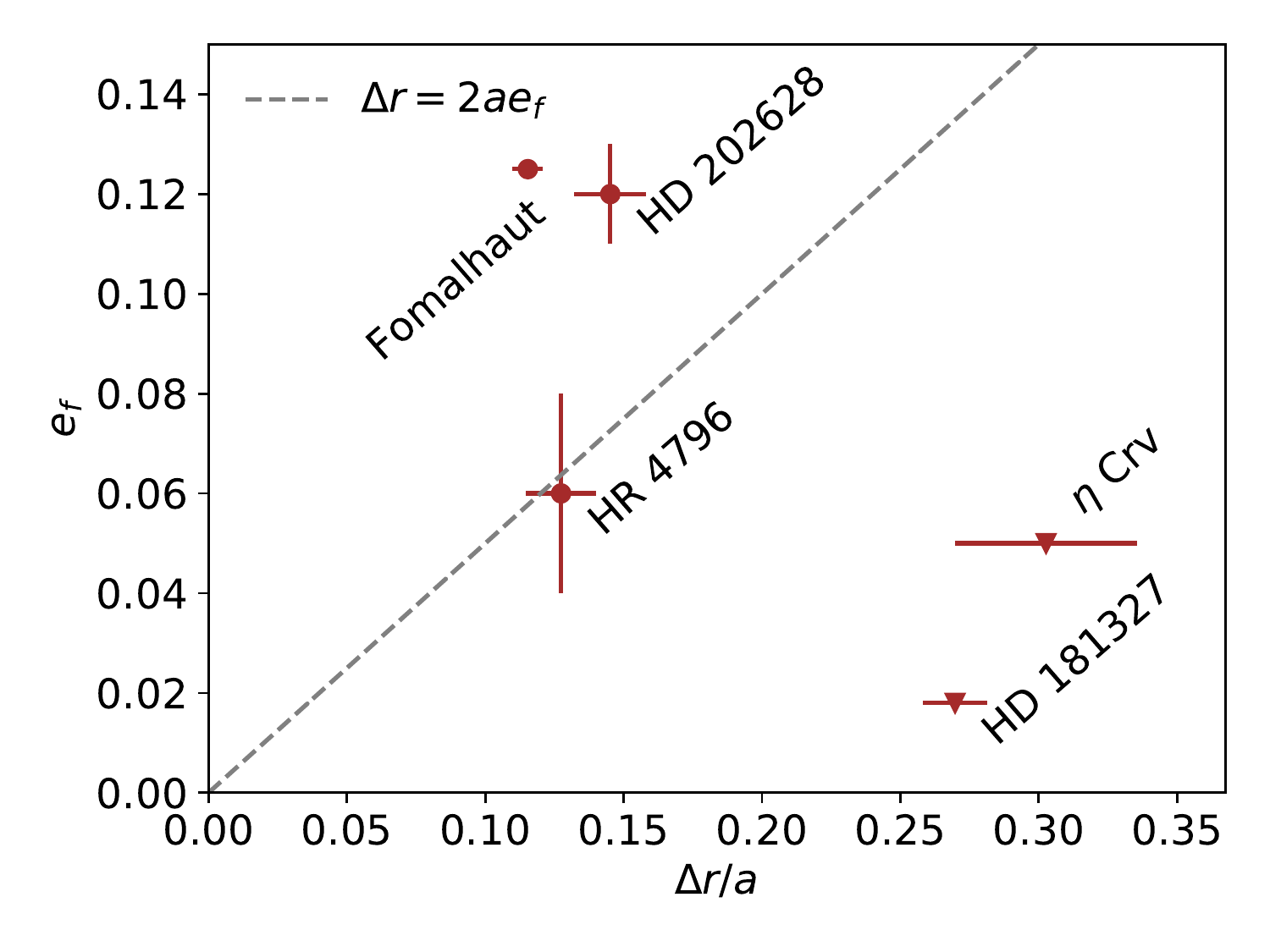}
  \caption{Forced eccentricity and relative ring width of selected
    debris discs. Widths are taken from modelling results
    \citep{2016MNRAS.460.2933M,2017MNRAS.465.2595M,2018MNRAS.475.4924K,2019AJ....158..162F},
    with the exception of Fomalhaut, which uses the FWHM of the narrower
    ansa (Figure \ref{fig:fom_rprof}). The line shows the expected width
    of the belt in a zero initial eccentricity secular perturbation
    scenario, showing that Fomalhaut and HD~202628's discs are narrower
    than expected, while HR~4796 is marginal (and like HD~202628 is not
    well resolved radially, so may be narrower). $\eta$ Corvi and
    HD~181327's discs are consistent with being circular, so their width
    originates from a range of semi-major axes, or eccentric bodies with
    randomly distributed pericenter directions.}
  \label{fig:ew}
\end{center}
\end{figure}

Having established that the Fomalhaut and HD~202628 debris rings are too
narrow to be explained by secularly perturbed particles that started
with near-zero initial eccentricities, several possibile origins are
discussed. Figure \ref{fig:ew} is shown to frame this discussion, which
shows the eccentricity and relative width of selected debris discs
\citep[see][for an extensive list]{2018ARA&A..56..541H}. Only discs with
ALMA-measured widths and eccentricities are shown, since these provide a
quantity that is not affected by small dust grain dynamics (the
eccentricity for HR~4796 was not detected by ALMA, so relies on
scattered light observation, but is included because the structure at
both wavelengths is consistent). While other discs have also been
observed with ALMA, these are wider than those shown, and are not known
to have significant eccentricity. The possible importance of Figure
\ref{fig:ew} is that the narrower discs are also those that are most
eccentric. Fomalhaut and HD~202628 both lie above the dashed line,
indicating that their width is narrower than expected given their forced
eccentricities. HR~4796 is a marginal case -- Kennedy \emph{et al.}
\cite{2018MNRAS.475.4924K} note that the disc is only marginally
resolved, and could be narrower (i.e. further to the left of the dashed
line).

Empirically, an important goal seems to be to fill out Figure
\ref{fig:ew}, in particular for discs whose widths are $\lesssim$30\% of
their semi-major axis; if the trend holds then it may be reasonably
expected that the mechanism that radially concentrates planetesimals
also excites a forced eccentricity. Figure \ref{fig:ew} may eventually
reveal a demarcation between narrow/eccentric and wide/circular disc
systems, which could indicate whether the concentration mechanism
operated in a given system.

The possible origins of narrow eccentric debris rings may be grouped
into two main types of scenarios, depending on whether the
$\sim$mm-sized dust observed by ALMA does, or does not, trace the
planetesimal population. In the former case the narrow rings are
explained by either modifying the secular perturbation scenario to
incorporate non-zero initial eccentricities, or perhaps discarding it
entirely, while in the latter case secular perturbations may be
retained, but the orbits of bodies changes as a function of their size
(i.e. particle free eccentricities "damp" towards the forced
eccentricity as they move down the collisional cascade). These
scenarios, plus several others, are discussed below.

\subsection{Eccentric initial conditions}

The simplest explanation for the narrow debris rings is that the
(perhaps naive) expectation of small initial eccentricities in the
secular perturbation scenario is not met. That is, the initial
conditions in \ewsp~may look more like the left two panels in Figure
\ref{fig:eg} than the upper right panel, though the initial
eccentricities need not be at the location of the forced
eccentricity. In this case there are two main scenarios to consider.

One possibility is that the planetesimals acquire significant coherent
eccentricities (i.e. with a preferred pericenter direction) prior to the
evolution that is assumed in a standard secular perturbation
scenario. For example, a single planet may exert some influence on
planetesimals prior to dispersal of the gas disc, but this influence is
diminished in a way that simply decreases the magnitude of the forced
eccentricity \citep[but with approximately the same forced pericenter
angle as after dispersal, perhaps similar to the scenario discussed
by][for circumbinary planetesimals]{2013ApJ...764L..16R}. Thus, when the
gas is removed the planetesimal eccentricities are already shifted
somewhat towards the forced eccentricity, and as they precess under the
influence of the same planet, they trace out a smaller circle in \ewsp,
resulting in a narrower debris ring. To also provide the tentative trend
in Figure \ref{fig:ew}, this process should also radially concentrate
planetesimals, perhaps by trapping them in a gas pressure maximum
exterior to the planet \citep[e.g.][]{2012A&A...545A..81P}, or trapping
dust in a narrow ring which then goes on to form said planetesimals via
the streaming instability \citep[e.g.][]{2009ApJ...704L..75J}.

It is of course possible that the planetesimals acquire \emph{all} of
their eccentricity before the gas disc is dispersed, in which case no
further eccentricity evolution via secular perturbations is
necessary. In this case, there is not necessarily a need for any planet,
as it may be that the initial eccentricitity excitation is related to
planet-free gas dynamics or dust/gas interaction, For example, Lyra \&
Kuchner \cite{2013Natur.499..184L} show that narrow eccentric rings
could form in planet-free discs with high dust/gas ratios, which could
occur in local gas pressure maxima, or more globally as the density
drops during gas disc dispersal.

How might evidence for these possible scenarios be sought? Most simply,
they both predict that planetesimals should form and acquire non-zero
eccentricites during the gas-rich phase of protoplanetary disc
evolution. One system showing a possible asymmetry is PDS~70, which
hosts a protoplanetary disc and at least one interior planet
\citep{2018A&A...617A..44K,2019NatAs...3..749H}. The dust component is
fairly narrow and shows both a brightness asymmetry \citep[the NW ansa
is brighter,][]{2018ApJ...858..112L}, and a stellocentric offset that
places this ansa farther from the star \citep[e.g. visible in Fig. 2
of][]{2019A&A...625A.118K}. While these combined properties might be
interpreted as apocenter glow \citep[e.g.][]{2016ApJ...832...81P}, the
disc is likely optically thick and the origin of the asymmetry may not
be so simple. Nevertheless, PDS~70 is a possible precursor to systems
such as Fomalhaut and HD~202628, and thus circumstantial evidence that
eccentric rings can exist before the debris phase.

A possible difference between the single and no-planet scenarios is that
in the presence of an eccentric planet the eccentricity of debris rings
should increase with time. Immediately after gas disc dispersal secular
perturbations have yet to act to their full effect, so debris rings only
posess their initial eccentricity, but later acquire their full
eccentricity, which is larger. A young disc such as that around HD~4796,
with eccentricity $\approx$0.06 \citep[i.e. less than the much older
systems Fomalhaut and HD~202628,][]{2017A&A...599A.108M} is suggestive,
but there are too few discs with well-constrained eccentricities and
widths for this test to be possible at present.

\subsection{Eccentricity damping}

While the prior discussion assumed that the observed mm-wave disc
structure is a true reflection of the underlying planetesimal orbits,
this assumption might be violated. Specifically, if particle free
eccentricities decrease as they become smaller, a debris ring could
appear narrower at mm wavelengths than the underlying planetesimal
belt. A possible way to damp orbital eccentricities in a debris disc is
in collisions, if the random velocities of post-collision fragments tend
to be smaller than the targets'. Because this damping decreases the
relative orbital velocities, the effect in an eccentric disc with a
preferred pericenter direction is to drive down the free
eccentricity. This scenario was explored analytically by Pan \&
Schlichting \cite{2012ApJ...747..113P}, and in general finds the
expected decrease in velocities for smaller objects.

This scenario however essentially relies on destructive collisions being
between objects of similar size, as the orbit of the center of mass has
a lower random velocity than either of the two bodies. If the typical
destructor is much smaller than the target body, then the post-collision
fragments will tend to retain their original velocities, and the damping
is inefficient.\footnote{Picture two people throwing pumpkins towards
  each other at high velocity, neither is likely to be splattered with
  fragments (most will fall to the ground below the collision).
  However, if one throws a much smaller object, such as a pebble, the
  pebble thrower will be hit by the fragmented pumpkin (which has much
  more momentum).} A significant size difference is generally expected
for destructive collisions, because the size distribution is always such
that smaller objects are much more common, and therefore the impactor
that destroys a larger object is usually the smallest one that can do so
\citep{2002MNRAS.334..589W}. In this case, orbital velocities do not
decrease significantly with object size, and the entire size
distribution inherits eccentricities from the largest bodies.

This picture lacks some nuance, and for example ignores the effect of
non-destructive collisions, but collisional damping seems unlikely to be
significant, and if anything eccentricities seem more likely to increase
in collisions. For example, in their simulations of impacts with
100km-diameter targets, Jutzi \emph{et al.} \cite{2019Icar..317..215J}
find that the largest fragments, which dominate the mass, tend to have
low velocities relative to the target center of mass, so the
approximation that all bodies share the same orbit as the collision
center of mass \citep[e.g.][]{2006A&A...455..509K} is reasonable.

A potentially attractive aspect for collisional damping is that it could
be observationally testable. If free eccentricities tend to decrease
with particle size in the mm to cm size regime, observations may be able
to measure different debris ring widths as a function of
wavelength. While the discussion above suggests that significant damping
is unlikely, modelling and observations focussing on this specific size
range would be valuable.

A side effect of damping is that the relative velocities between smaller
particles is also smaller, leading to longer collisional lifetimes and a
steeper size distribution than would be expected if all objects share
the same random velocities \citep{2012ApJ...747..113P}. The steeper size
distribution in turn means that for a given observed disc brightness
(which is proportional to the surface area of small grains), the
inferred total mass (which is dominated by large bodies) can be much
smaller. Damping therefore provides a possible resolution of the debris
disc ``mass problem'' \citep[see][for discussion of this
issue]{2018MNRAS.474.2564K}, in which disc masses in systems such as
HR~4796 are inferred to be implausibly large
\citep[e.g.][]{2018MNRAS.475.4924K}.

\subsection{A single event}

A third way to explain the narrow debris rings is to discard the
standard debris disc paradigm entirely (i.e. the idea of dust derived
from an underlying population of planetesimals undergoing continuous
collisions in a pseudo-steady state). That is, to consider transient
and/or stochastic events as a way to produce a population of objects on
similar eccentric orbits. The scenario for creating a narrow debris ring
is then based on the breakup of a single large body, which must have
been destroyed in such a way that the velocity dispersion of fragments
is small (i.e. similar to described above). The scenario may therefore
be very similar to that explored by Jackson \emph{et al.}
\cite{2014MNRAS.440.3757J}, where debris is released from a large body
in a collision. This scenario tends to produce non-axisymmetric dust
density distributions, as more dust is concentrated at the spatial
location of the original dust production event (and more radially
distributed on the opposite side of the star). However, for a small
velocity dispersion the asymmetry will be smaller, and if there are
other perturbing bodies in the system, the asymmetry is decreased as the
fragments' orbits precess.

In some ways this scenario is similar to a normal debris ring scenario,
in that once created, a population of fragments evolves in the same way,
and is indistinguishable from a planetesimal population with similar
orbits. It is therefore very hard to rule out that narrow debris rings
(eccentric or otherwise) originate from planetesimals that are
themselves a family of fragments.\footnote{Another difference noted by
  Cataldi \emph{et al.} \cite{2019arXiv190407215C} is that if CO gas is
  also liberated by the collisional evolution, then very different
  levels of atomic carbon (which results from rapid photodissociation of
  CO) may result. As carbon is not necessarily easily removed from the
  system, it builds up over time, so a system where the observed debris
  is the result of a more recent event should have lower levels of
  carbon than one that has been evolving for longer.} A basic objection
however is why only a single debris ring should be detected around a
given star, and not a series of near-concentric rings that arise from
similar events. A more serious issue is that inferred masses for debris
discs can be of order tens of Earth masses \citep[e.g. 20-30$M_\oplus$
for Fomalhaut,][]{2002MNRAS.334..589W}, and it therefore seems unlikely
that this mass in solid fragments can be created from a single event.

\subsection{Shepherding panets}

Boley \emph{et al.} \cite{2012ApJ...750L..21B} proposed that the width
of the Fomalhaut debris ring is constrained by a pair of shepherding
planets, by analogy with Uranus' $\epsilon$ ring and Saturn's F ring. In
these cases the moons were inferred to exist prior to their discovery
because the rings should rapidly spread radially, and some external
force is required to confine them \citep{1979Natur.277...97G}. As
discussed above, collisions in debris discs are such that significant
spreading of narrow rings is not necessarily expected. This picture does
not however mean that shepherding moons do not exist, just that they are
not required by the analogy that narrow debris rings appear similar to
narrow planetary rings.

As discussed at the outset, a problem with planet-interaction scenarios
is that the putative planets are often not detectable. In the case of
Fomalhaut, the inferred shepherding planet masses were several Earth
masses \citep{2012ApJ...750L..21B}, meaning that their detection at
$\sim$140\,au from the star is currently impossible. An avenue that has
not yet been explored is whether shepherding planets should induce
detectable azimuthal structure in the ring edges near the planets
\citep{1984NASCP2330..609D}; such a study would require simulations, and
no doubt higher spatial resolution imaging. In the meantime, the
shepherding planet scenario is considered more complicated than is
required by the data; while shepherding requires two eccentric planets,
the secular perturbation scenario only requires one.

\subsection{Massive debris ring}

The secular perturbation theory as applied to debris discs usually
assumes that the belt mass can be ignored; the planet perturbs the belt,
but the belt does not perturb the planet (i.e. cause it to precess). In
the case of eccentric debris rings this assumption seems justified, as a
precessing planet would cause the forced eccentricity to move in
\ewsp~over time, and differing precession rates for ring particles at
different semi-major axes would cause the ring to quickly lose the
observed coherence. However, this issue might be circumvented with
shepherding satellites as discussed above, or if the ring has sufficient
mass that self-gravity forces all particle pericenters to remain
similar. If such a coherence-maintaining mechanism could operate, then
it is possible that the debris ring was initially near-circular, and
gained an observable eccentricity through mutual interaction with an
initially eccentric planet. While these ideas were developed for
planetary rings in the Solar system \cite[e.g.][]{1979Natur.277...97G},
whether they apply here is not clear, so this scenario requires further
work.

\section{Conclusions}\label{s:conc}

This paper shows that the eccentric Fomalhaut and HD~202628 debris rings
are narrower than expected, based on a secular perturbation model and
the expectation of near-zero initial eccentricities. In the case of
Fomalhaut, this narrowness is clear from simply measuring the radial
profile as observed by ALMA (Figure \ref{fig:fom_rprof}), but for
HD~202628 the s/n is per beam is lower so the conclusion drawn from
several similar eccentric ring models.

What does this narrowness mean? The most likely implication is that in a
secular perturbation scenario the planetesimals did not initially have
near-circular orbits. This prior excitation could be a consequence of
planetary perturbations within the gas-rich protoplanetary disc, but if
other processes can produce coherently eccentric planetesimal orbits,
may not require a planet at all. In either case however, a clear
prediction is that eccentric rings should exist within protoplanetary
discs. Whether these rings are observable is less clear, but PDS~70,
host to a protoplanetary disc that shows both geometric and brightness
asymmetry, and at least one planet, is singled out as a possible
progenitor of systems such as Fomalhaut and HD~202628. Making and
extending such links would be valuable to understand whether debris
discs or their direct progenitors exist within gas-rich protoplanetary
discs.

An alternative possibility is that the planetesimal belt is wider than
observed with ALMA, but that objects are damped as they fragment to
smaller sizes. This possibility is disfavoured, but may be testable if
mm to cm-size grains have different enough eccentricities. Observational
tests of collisional damping may also be relevant to the debris disc
mass problem.

A plot of belt forced eccentricity against relative radial width (Figure
\ref{fig:ew}) provides circumstantial evidence of a trend that the
narrowest debris rings are also the most eccentric. However, a lack of
eccentric systems limits the numbers in this Figure, and more systems
will need to be characterised to explore this possible relation
further. For a start, HR~4796 is identified as possibly being narrower
than expected, and should be imaged at higher spatial resolution.

\vskip1pc


\dataccess{This paper makes use of the following ALMA data:
  ADS/JAO.ALMA\#2015.1.00966.S, ADS/JAO.ALMA\#2016.1.00515.S, which are
  available from the ALMA archive:
  \href{http://almascience.nrao.edu/aq/}{http://almascience.nrao.edu/aq/}.
  Relevant code for this research work is stored in GitHub:
  \href{https://github.com/drgmk/eccentric-width}{https://github.com/drgmk/eccentric-width}
  and have been archived within the Zenodo repository:
  \href{https://doi.org/10.5281/zenodo.3832148}{https://doi.org/10.5281/zenodo.3832148}.}


\competing{GMK declares no competing interests.}

\funding{GMK is supported by the Royal Society as a Royal Society
  University Research Fellow.}

\ack{Thanks to both referees for constructive reports, Jane Huang for
  noting the possibly eccentric structure of the PDS~70 disc, to Luca
  Matr\`a for advice/discussions on modelling ALMA data, to Meredith
  MacGregor and Virginie Faramaz for discussions on their prior work on
  Fomalhaut and HD~202628, and to Mark Wyatt for the reminder that
  planetesimal belts have mass. ALMA is a partnership of ESO
  (representing its member states), NSF (USA) and NINS (Japan), together
  with NRC (Canada), MOST and ASIAA (Taiwan), and KASI (Republic of
  Korea), in cooperation with the Republic of Chile. The Joint ALMA
  Observatory is operated by ESO, AUI/NRAO and NAOJ.}


\pagebreak



\appendix

\section{Posterior distributions for model fitting}

\begin{figure*}
  \includegraphics[width=1\textwidth]{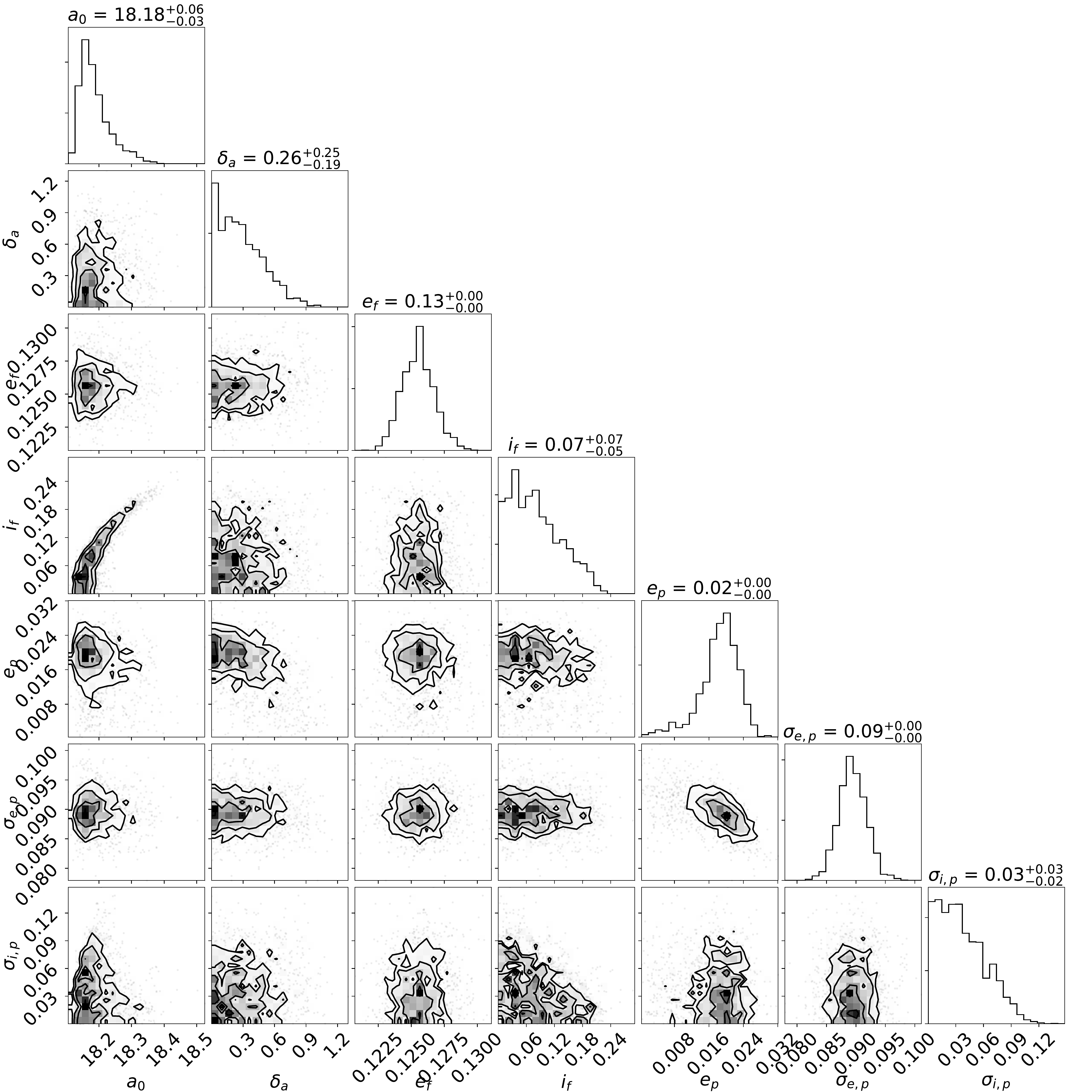}
  \caption{Posterior parameter distributions for the full model of the
    Fomalhaut ring with a uniform distribution of semi-major axes. Only
    parameters related to the eccentric disc model are shown here. Each
    off-diagonal panel shows the 2-d distributions, with 1, 2, and
    3$\sigma$ contours in regions of high density, and individual dots
    in regions of low density. The diagonal panels show the 1-d
    distributions.}
    \label{fig:fom-da-full}
\end{figure*}

\begin{figure*}
  \begin{center}
    \includegraphics[width=0.6\textwidth]{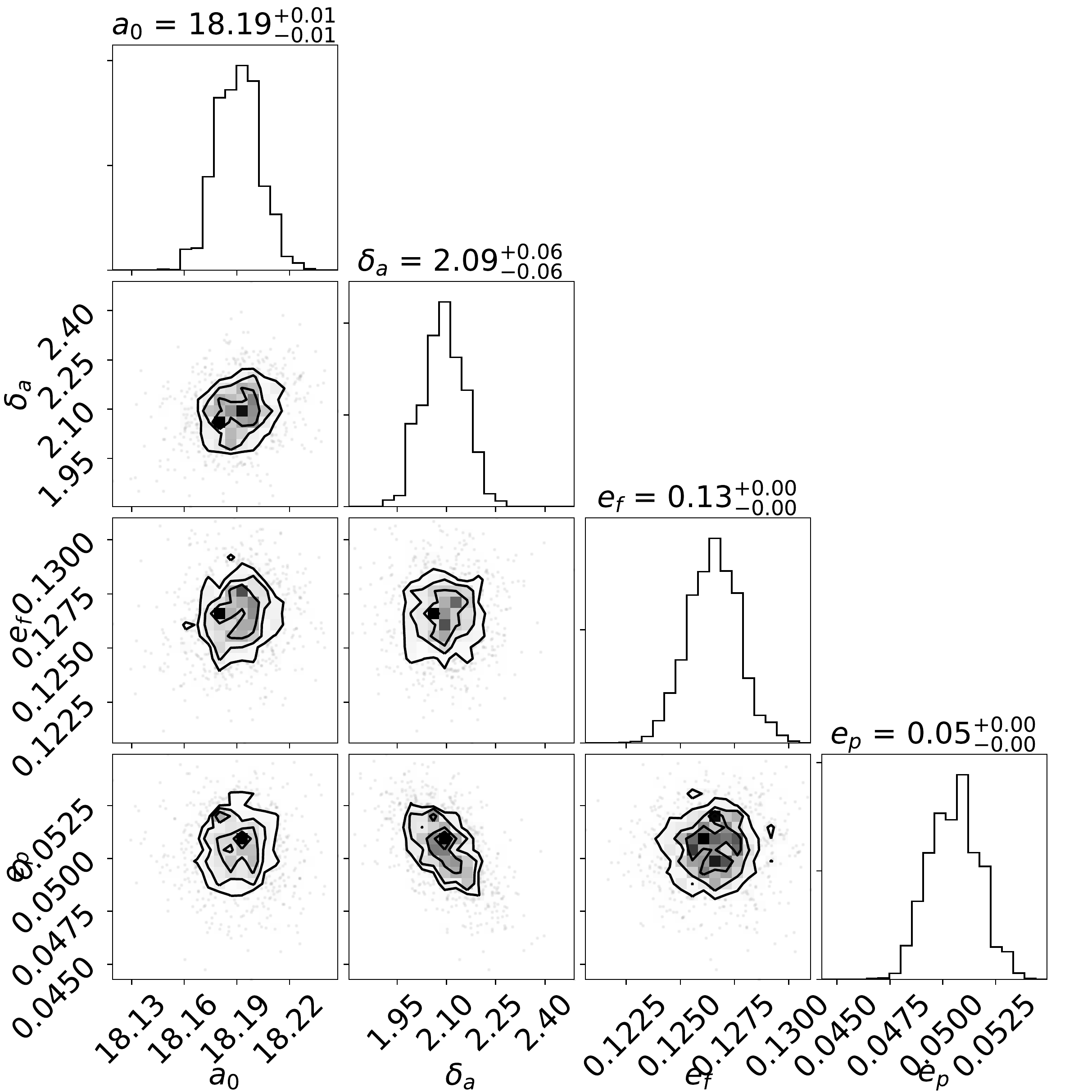}
    \caption{Posterior parameter distributions for the simple model of
      the Fomalhaut ring with a uniform range of semi-major axes. Only
      parameters related to the eccentric disc model are shown
      here. Each off-diagonal panel shows the 2-d distributions, with 1,
      2, and 3$\sigma$ contours in regions of high density, and
      individual dots in regions of low density. The diagonal panels
      show the 1-d distributions.}
    \label{fig:fom-da-flat}
  \end{center}
\end{figure*}

\begin{figure*}
  \includegraphics[width=1\textwidth]{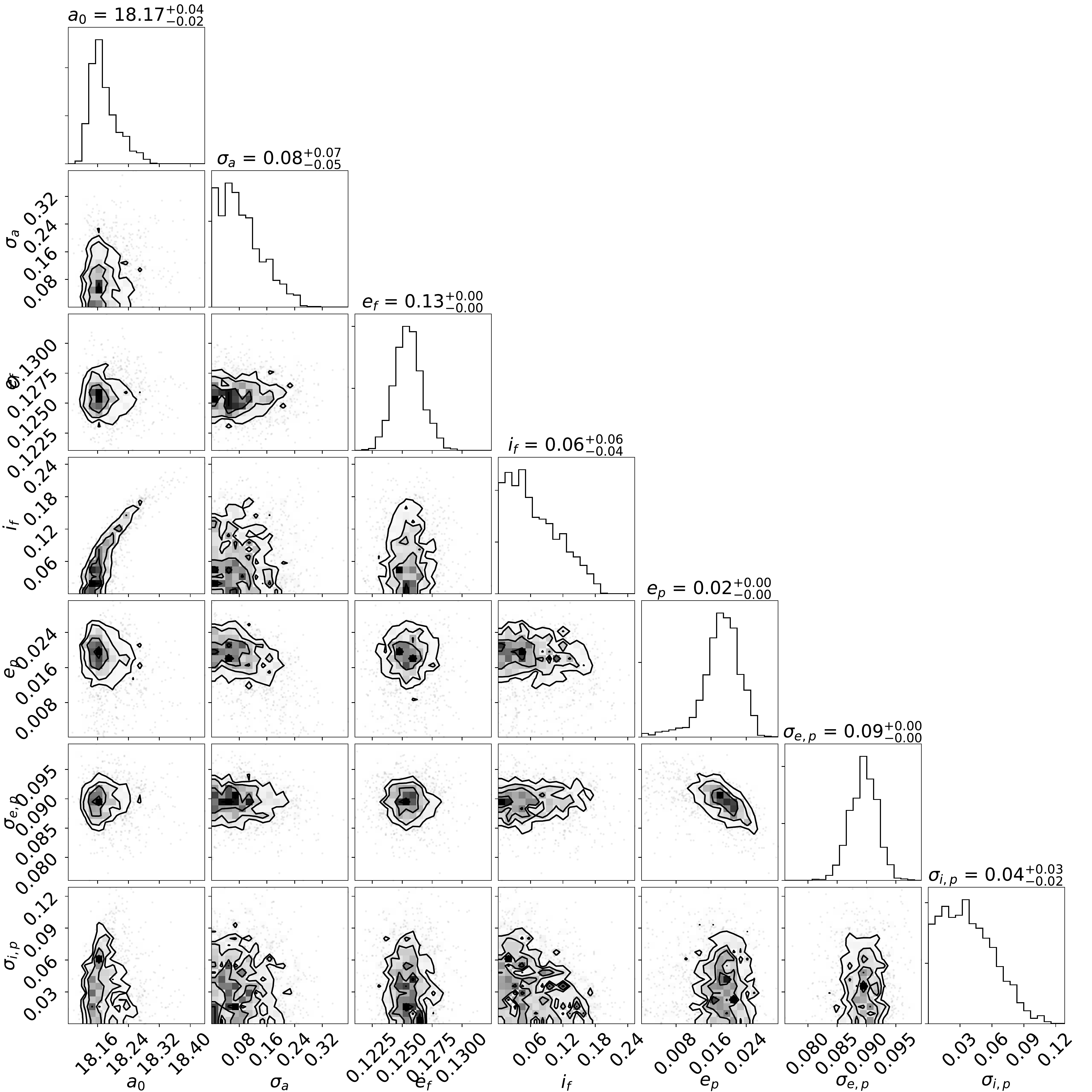}
  \caption{Posterior parameter distributions for the full model of the
    Fomalhaut ring with a Gaussian range of semi-major axes. Only
    parameters related to the eccentric disc model are shown here. Each
    off-diagonal panel shows the 2-d distributions, with 1, 2, and
    3$\sigma$ contours in regions of high density, and individual dots
    in regions of low density. The diagonal panels show the 1-d
    distributions.}
    \label{fig:fom-g-full}
\end{figure*}

\begin{figure*}
  \includegraphics[width=1\textwidth]{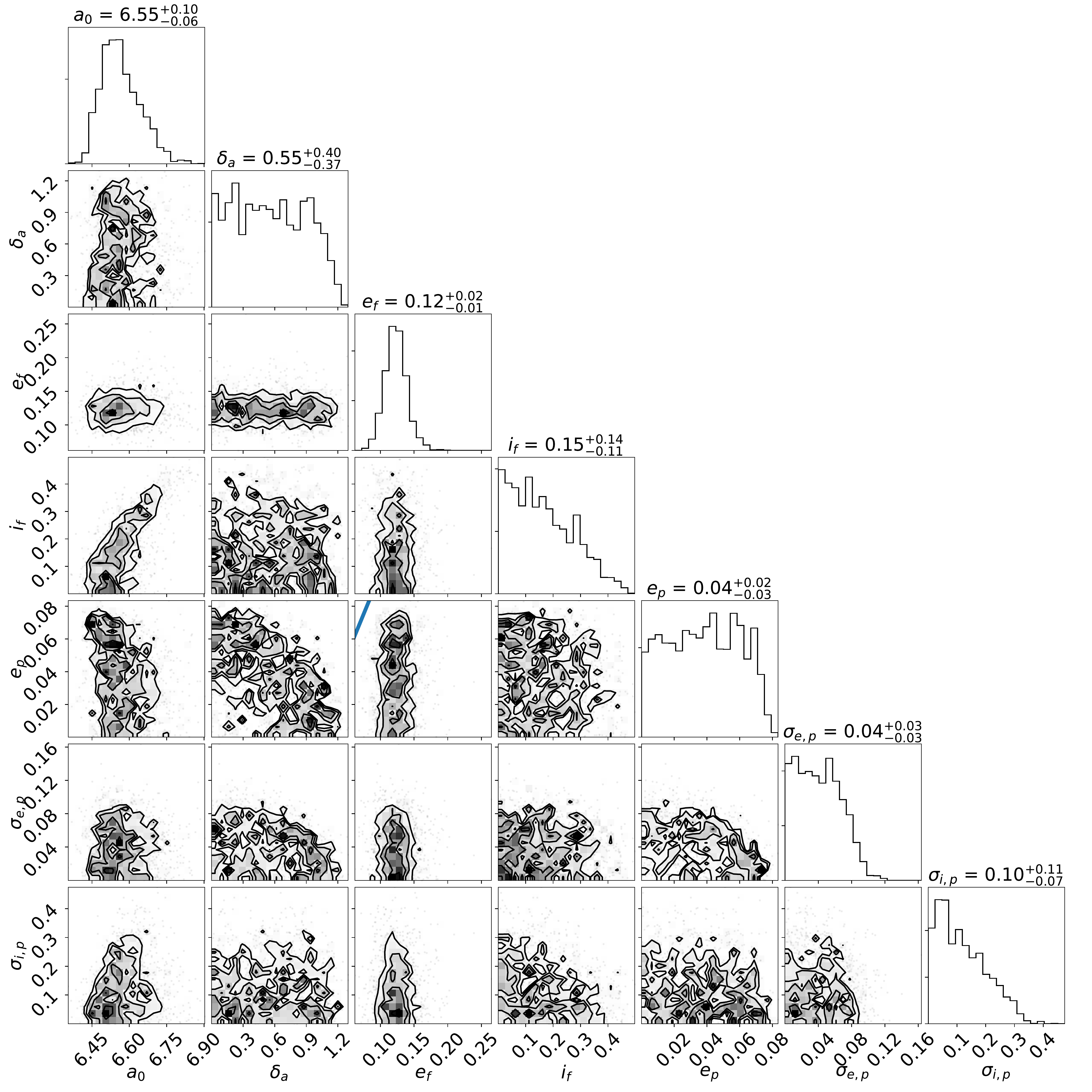}
  \caption{Posterior parameter distributions for the full model of the
    HD~202628 ring with a uniform range of semi-major axes. A line with
    $e_p=e_f$ is shown in the appropriate panel, illustrating that the
    proper eccentricity is significantly lower than the forced
    eccentricity. Only parameters related to the eccentric disc model
    are shown here. Each off-diagonal panel shows the 2-d distributions,
    with 1, 2, and 3$\sigma$ contours in regions of high density, and
    individual dots in regions of low density. The diagonal panels show
    the 1-d distributions.}
    \label{fig:hd-da-full}
\end{figure*}

\begin{figure*}
  \begin{center}
    \includegraphics[width=0.6\textwidth]{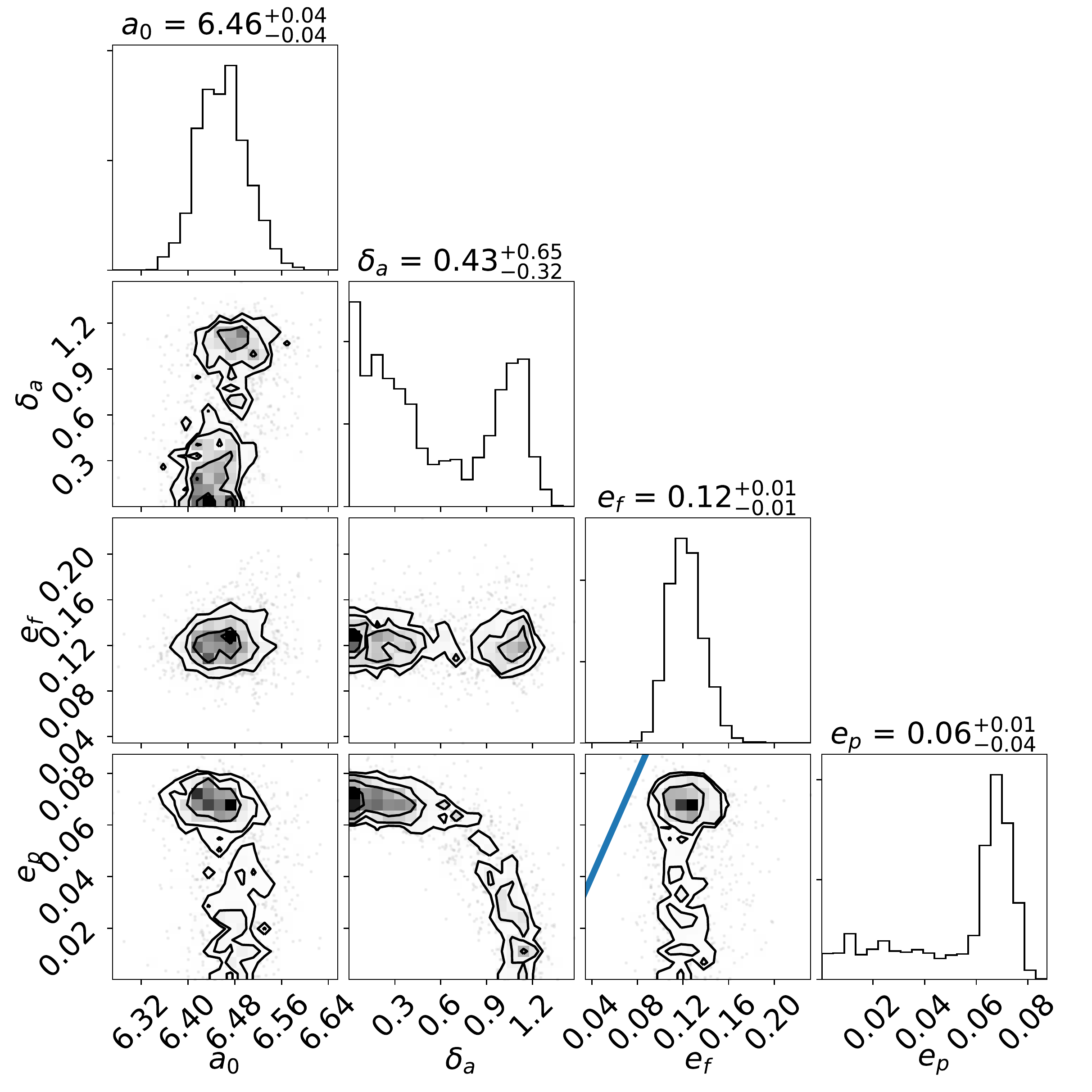}
    \caption{Posterior parameter distributions for the simple model of
      the HD~202628 ring with a uniform range of semi-major axes. A line
      with $e_p=e_f$ is shown in the appropriate panel, illustrating
      that the proper eccentricity is significantly lower than the
      forced eccentricity. Only parameters related to the eccentric disc
      model are shown here. Each off-diagonal panel shows the 2-d
      distributions, with 1, 2, and 3$\sigma$ contours in regions of
      high density, and individual dots in regions of low density. The
      diagonal panels show the 1-d distributions.}
    \label{fig:hd-da-flat}
  \end{center}
\end{figure*}

\begin{figure*}
  \includegraphics[width=1\textwidth]{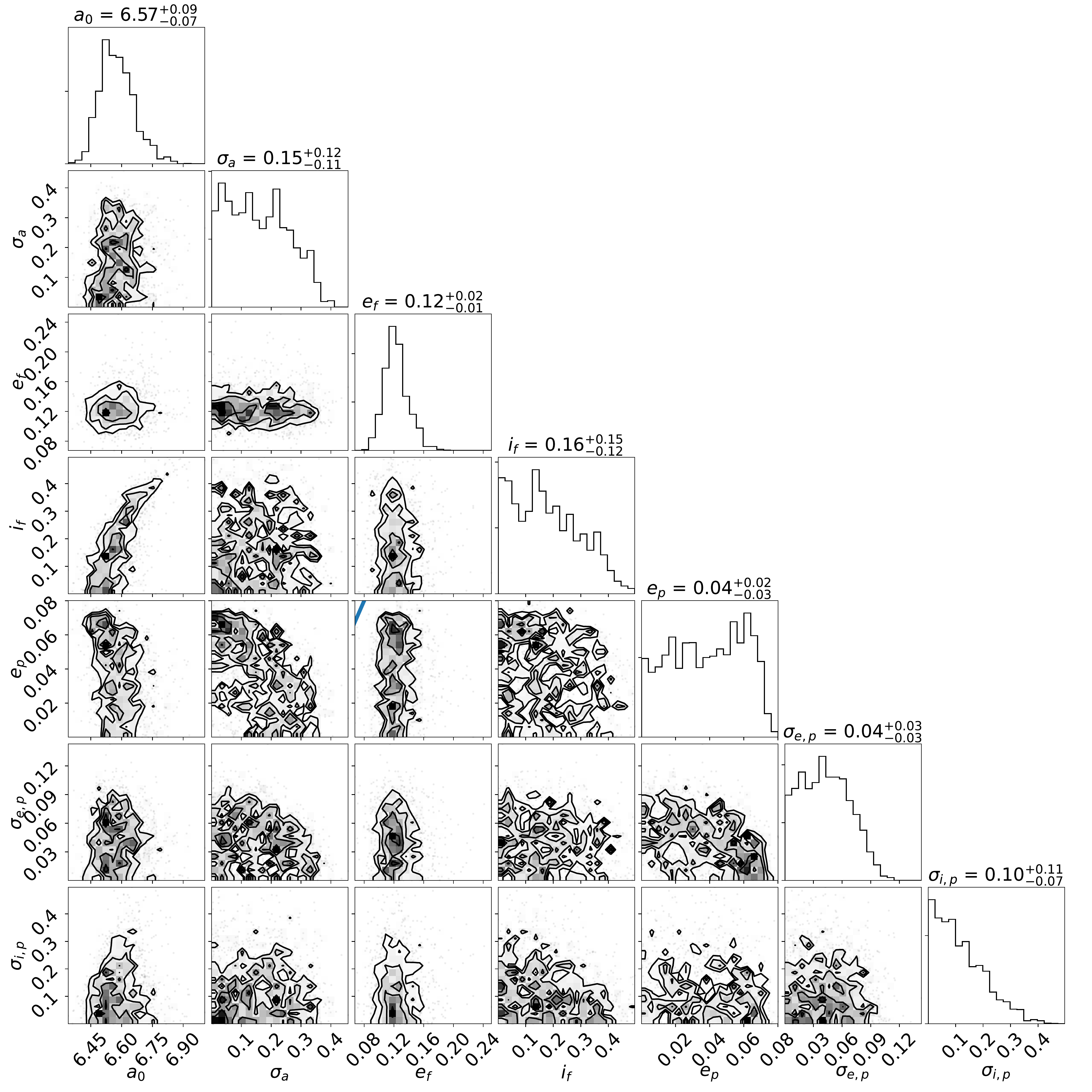}
  \caption{Posterior parameter distributions for the full model of the
    HD~202628 ring with a Gaussian range of semi-major axes. A line with
    $e_p=e_f$ is shown in the appropriate panel, illustrating that the
    proper eccentricity is significantly lower than the forced
    eccentricity. Only parameters related to the eccentric disc model
    are shown here. Each off-diagonal panel shows the 2-d distributions,
    with 1, 2, and 3$\sigma$ contours in regions of high density, and
    individual dots in regions of low density. The diagonal panels show
    the 1-d distributions.}
    \label{fig:hd-g-full}
\end{figure*}

\end{document}